\documentclass [12pt,preprint]{aastex}

\usepackage{graphicx}  
\usepackage{eurosym}
\usepackage{amsfonts}
\usepackage{amsmath}

\usepackage{longtable}

\def\kms{km s$^{-1}$}

\shorttitle{Periodicity in  NGC 5548}

\shortauthors{Bon et al.}

\begin{document}

\title{Evidence for periodicity in 43-year-long monitoring of NGC 5548
}

\author{E. Bon\altaffilmark{1,2},
S. Zucker\altaffilmark{3},
H. Netzer\altaffilmark{4},
P. Marziani\altaffilmark{5},
N. Bon\altaffilmark{1,2},
P. Jovanovi\'{c}\altaffilmark{1,2},
A. I. Shapovalova\altaffilmark{6},
S. Komossa\altaffilmark{7},
C. M. Gaskell\altaffilmark{8},
L. \v C. Popovi\'c\altaffilmark{1,2},
S. Britzen\altaffilmark{7},
V. H. Chavushyan\altaffilmark{9},
A. N. Burenkov\altaffilmark{6},
S. Sergeev\altaffilmark{10},
G. La Mura\altaffilmark{11},
J. R. Vald\'es\altaffilmark{9} \&
M. Stalevski\altaffilmark{1,12,13}}

\altaffiltext{1}{Astronomical Observatory, Volgina 7, 11060 Belgrade, Serbia;}
\altaffiltext{2}{Isaac Newton Institute of Chile, Yugoslavia branch, Serbia;}
\altaffiltext{3}{Department of Geosciences, Tel-Aviv University, Tel-Aviv 6997801, Israel;}
\altaffiltext{4}{School of Physics and Astronomy and the Wise Observatory,
The Raymond and Beverly Sackler Faculty of Exact Sciences, Tel-Aviv
University, Tel-Aviv 6997801, Israel;}
\altaffiltext{5}{INAF, Osservatorio Astronomico di Padova, Padova, Italia;}
\altaffiltext{6}{Special Astrophysical Observatory of the Russian AS, Nizhnij Arkhyz, Karachaevo-Cherkesia 369167, Russia;}
\altaffiltext{7}{Max-Planck-Institut f{\"u}r Radioastronomie, Auf dem H{\"u}gel 69, 53121 Bonn, Germany}
\altaffiltext{8}{Department of Astronomy and Astrophysics, University of California at Santa Cruz, Santa Cruz, CA 95064.}
\altaffiltext{9}{Instituto Nacional de Astrofísica, \'Optica y Electr\'onica,
Apartado Postal 51, CP 72000, Puebla, Pue, Mexico, Mexico;}
\altaffiltext{10} {Crimean Astrophysical Observatory, P/O Nauchny, Republic of Crimea
298409, Russia;}
\altaffiltext{11}{Dipartimento di Fisica e Astronomia ``G. Galilei",
Universit\`a degli Studi di Padova,
Vicolo dell'Osservatorio 3, 35122 - Padova, Italy}
\altaffiltext{12}{Departamento de Astronom\'\i a, Universidad de Chile, Camino El Observatorio 1515, Casilla 36-D Santiago, Chile}
\altaffiltext{13}{Sterrenkundig Observatorium, Universiteit Gent, Krijgslaan 281-S9, Gent, 9000, Belgium}

\begin{abstract}

{We present an analysis of 43 years (1972 to 2015) of spectroscopic observations of the Seyfert 1 galaxy 
NGC 5548. This includes 12 years of new unpublished observations (2003 to 2015). 
We compiled about 1600 H$\beta$  spectra and analyzed the long-term spectral variations 
of the 5100 {\AA} continuum and the H$\beta$ line. Our analysis is based on standard procedures, 
including the Lomb–Scargle method, which is known to be rather limited to such heterogeneous 
data sets, and a new method developed specifically for this project that is more robust and 
reveals a $\sim$5700 day periodicity in the continuum light curve, the H$\beta$ light curve, and the 
radial velocity curve of the red wing of the H$\beta$ line. The data are consistent with orbital 
motion inside the broad emission line region of the source. We discuss several possible mechanisms 
that can explain this periodicity, including orbiting dusty and dust-free clouds, 
a binary black hole system, tidal disruption events, and the effect of 
an orbiting star periodically passing through an accretion disk.}

\end{abstract}


\section{Introduction}

Despite {much progress in recent years, many
fundamental questions about the structure and kinematics of the innermost material in active
galactic nuclei (AGNs) remain unanswered.
Two key features of thermal AGNs are (i) that the IR to X-ray continuum is highly variable and
(ii) that they have a broad-line region (BLR).
Because the wavelengths of emission lines are well known the effective line-of-sight
velocity of line-emitting gas is known.
Line profiles of broad lines are thus an important constraint on models of the inner regions
of AGNs.  Furthermore, the ratios of intensities of
different lines depend on the physical conditions of the environment of the gas such as
the density and radiation field.}

Variability of {AGNs} on short and long timescales
{potentially provides} valuable insights about the physics of accretion, mechanisms of fueling
nuclei and the growth of supermassive black holes.
{Because the BLR gas is close to the center of the AGN, it readily responds
to continuum variability.  Cross correlating the
variability of broad lines with variability of the continuum readily gives the sizes of
line-emitting regions \citep{CherepashchukLyutyi73, GaskellSparke86}.
In addition, the velocity-dependence of a line's response to
continuum variability provides information about the kinematics and dynamics of the line emitting gas
\citep[see {e.g.,}][]{Gaskell88, Maoz94, Netzer97, Peterson97, Netzer2013}.}

{Because of its brightness,
NGC 5548 was among the first Seyfert galaxies to be studied.
It was the first galaxy to have optical variability reported \citep{Deich66}.
Between 1966 \citep{Dibai68} and 1970-71 \citep{Anderson71, Ulrich72}
there was a large change in the profile of the broad H$\alpha$ line.
The optical continuum was also highly variable \citep{Lyutyi73}.
The first reverberation mapping of NGC 5548 \citep{PetersonGaskell86}
showed that H$\beta$ responded to continuum changes with a delay
of only a few weeks.   Because its brightness, reliable variability, and
convenient broad-line region (BLR) size, NGC 5548 was recognized as
an easy target for reverberation mapping.  It has therefore been the
subject of much monitoring for several decades  \citep[see e.g.,][ and references
within]{PetersonGaskell86, Netzer90, Koratkar91, Clavel91,
Peterson91, Peterson92, Dietrich93, Korista95, Peterson99, Kaspi2000, Dietrich01,
Peterson02, Shap04, Sergeev07, Bentz07, Popovic2008, Bentz2009,
Denney09, Denney10, DeRosa15, Li2016}.} Studies at radio, visible,
{UV,} and X-ray wavelengths indicate violent processes including
ejection of gas, \citep[{e.g.,}][]{Kollatschny13a, Kaastra2014}.

According to \cite{Sergeev07}{,} inspection of individual broad H$\beta$
profiles over a {30-year} period reveals that the broad emission line
profiles can
undergo dramatic changes (from a typical single-peaked profile centered
near the systemic redshift of the galaxy, to profiles that show
prominent blue or red peaks. Descriptions of blue and red peaks are
presented in
many papers {\citep[see
e.g.][]{Anderson71, Ulrich72, Ptak73, Peterson87, Sergeev92, Shap04, Shap06, Popovic2008, Li2016}.}

AGN variability  is detected at {essentially} all wavelengths \citep{Netzer2013}
and {on all} time scales \citep[the light crossing time of the system, the rotational
period of the central power-house with the associated line emitting gas, and the
viscous time of the central accretion disk, 
see][]{Czerny2006, Netzer2013}. The light crossing time of the broad line
region (BLR) of more than 60 sources has been used to characterize the dimension of the
system (R$_{BLR}$) and to estimate black hole (BH) mass \citep[see e.g.][and references
therein]{Gask03, Kaspi2000, Bentz13, Du15}. While most AGN vary by a factor
of a few in the optical band, there have been a few examples  with systematic
long-term trends, like Mrk 590, which shows: a) an overall long-term decrease  by a factor
100, along with b) a change in Seyfert type \citep{Denney14}, suggesting a significant
decrease in accretion rate. Despite many searches for semi-periodic variations in the
AGN lightcurves, few {convincing} candidates have been found so far\citep[see
e.g.][]{Sillanpaa88, Lehto1996, Fan1998, Rieger2000, Valtaoja2000, DePaolis2003,
Sudou2003, Guo2006, Gezari07, Guo2014, Liu15, Bon12, Graham15a, Graham15b, Shap16}.

Periodic variations could be produced
various ways, including binary black hole (BH) systems,
tidal disruption events (TDE) and more
\citep[e.g.] [and references therein]
{Gaskell83, Sillanpaa88, Komossa06, Bog08, Gask09, erac2012,
Pop2012, Bon12, Valtonen12, Bog15, KomossaZensus2015, Komossa15}.
Distinguishing between scenarios {requires} extremely
long monitoring,
{that is} only available for a handful of sources.
While AGN variability  has been
documented for many decades, {only a} few light curves span a time interval as long as
100 years [e.g., NGC 4151, from 1906  \citep[][]{Oknyanskij2007},  3C273, from
the 1880s  \citep[][]{SmitHoff1963}, and OJ287, from 1891
\citep{ValtonenSila2011} to the present.
Therefore, well-covered long-term lightcurves
of nearby AGN are required to search for the
presence or absence of periodic signatures.
This paper presents the analysis of very long
duration light curves of NGC 5548
that span over 43 years and 1600 optical spectra,
including 12 years of new data.
The aim of this paper is to search for periodicity
in the continuum light curve, the emission-line light curves, and the radial
velocity curves. The structure of our paper is as follows. In
\textbf{\S\ref{observation}} we present information about the new
observations. In section \textbf{\S\ref{methods}} we explain our
methods of calibrations in section \textbf{\S\ref{results}}
light and radial velocity curves.
Various {possible interpretations are} given in section \textbf{\S\ref{interpretation}}.
Finally, in \textbf{\S \ref{conclusions}} we summarize
our results and present the conclusions.

\section{Observations and data reduction}\label{observation}

We analyzed over 1600 spectra of NGC 5548 in the H$\beta$\ spectral interval, { covering over 43} years:
 (a) {archival spectra {obtained by K.~K.~Chuvaev} from 1972-1988 \citep{Sergeev07} prior to the {\it International AGN
Watch} (IAW) campaigns.  These early spectra} were recorded on photographic plates acquired with an image tube at the 2.6 m Shajn Telescope
of the Crimean Astrophysical Observatory.
 (b) {the intensive, 13-year study from 1988 to 2002 of the {IAW} program \citep{Peterson02}.  This provided 1530
optical continuum measurements and 1248 H$\beta$ measurement}\footnote{The {IAW} data 
could be obtained in the digital format from the following link:
http://www.astronomy.ohio-state.edu/~agnwatch/data.html}.
 (c) {a spectral monitoring program with the 6-m and 1-m
telescopes of the Special Astrophysical Observatory (SAO) in Russia from 1996 to 2002, and the 2.1 m-telescope of Guillermo Haro
Observatory (GHO) in Cananea, Mexico from 1996 to 2003 \citep{Shap04}.}
 (d) {more recent, }unpublished observations of the same program covering {2003} - 2013,
observed at SAO (see Table \ref{Tab_SAO}) and a continuation of { the} monitoring campaign presented in \cite{Shap04},
 (e) {spectra from the new IAW campaign obtained at Asiago observatory in 2012, 2013 and 2015\footnote{with kind permission of PI
Bradley Peterson to use IAW data published in \cite{Peterson2013} and unpublished data from 2012, 2013 and 2015, observed at Asiago, as
there is no publication based on those data yet.
}
 }
and (f) new unpublished observation from 2013 from { the} Asiago observatory ({also given in}
Table \ref{Tab_SAO}).

{Details of the additional} optical spectra { obtained at} INOAE and Asiago {are as follows.  The} SAO
and INOAE spectra were obtained with the {6-m and 1-m} telescopes { at SAO and with the INAOE 2.1-m} telescope
at the Guillermo Haro Observatory (GHO) at Cananea, Sonora, Mexico.  { In all cases observations were made with long-slit}
spectrographs equipped with CCDs.
The typical wavelength range covered was from 4000 \AA\  to 7500 \AA,
the spectral resolution was 4.5-15 \AA, and the S/N ratio was $>$ 50
in the continuum near H$\alpha $ and H$\beta$. Spectrophotometric standard
stars were observed every night. The logs of these new observation are presented
in the Table \ref{Tab_SAO}.
{We also} include a set of unpublished spectra observed at the {1.22-m} telescope of the Asiago
Astrophysical Observatory, configured in long-slit
spectroscopy mode. The total exposure time was 3600s, divided in multiple runs
of 600s or 1200s each, in order to prevent saturation of the strongest
emission lines. The spectrograph used a 300 { lines}/mm grating with a {300 $\mu$m} slit width, achieving a
spectral resolution {$R \simeq$ 600} between
3700~\AA\ and 7500~\AA. {W}avelength calibration was obtained {using} FeAr
comparison lamps, while the flux calibration was performed with the
observation of the spectro-photometric standard stars
{Feige 34 and Feige 98}. Cosmic rays were identified
and masked out through the combination of the different short exposures.

\section{Methods of analysis}\label{methods}

 {With the goal to analyse NGC 5548 spectra from 43 years of monitoring campaigns, we performed full spectrum fitting analysis using}
ULySS code \citep{Kol09}\footnote{{The ULySS full spectrum fitting package is available at: http://ulyss.univ-lyon1.fr/}} {
that we} adopted {for fitting} Sy1 spectra with {models representing a} linear combination
of non-linear model components {--} emission lines, Fe II templates, AGN continuum and { the }stellar
population { of the host galaxy}. {This is the first time the package has been used to analyze spectra of broad lines in AGNs,
{while before it was used for 
(i) determining stellar
atmospheric parameters using the models of stellar atmosphere \citep{Wu11} and (ii) studying the history of stellar populations
 \citep{Bouch10,Kol11,Kol13}}. Recently, \cite{Bon2014} tested the accuracy of the code in recovering stellar population and gas parameters
in Type 2 AGNs. }

We obtained light curves and radial velocity curves for all {spectra
and} searched them for possible periodicities. We used standard methods for {treating}
unevenly { spaced} data, and {a} new method specially developed for this purpose
tailored for our specific special conditions of very few cycles  and very
specific sampling characteristics of here obtained data series
(see section \ref{lc-analysis}).

\subsection{Line and continuum fittings}\label{FM1}

{Since ULySS gives us a choice of defining and including components, we adjusted it to analyze simultaneously all components
that
contribute to the flux in the wavelength region around H$\beta$.  For
analyzing variability of NGC 5548} we defined the {model, $M(x)$,} as follows:
\begin{equation}
\begin{array}{r}
M(x) = P(x) \ ([T (x) \otimes G(x)] + C(x) + N(x) + \\
B(x) + \sum_{i=1}^{4} FeII_{i}(x) +
\sum_{j=1}^{n} S_{j}(x)),
\end{array}
\end{equation} \label{eq:1}
where {$M(x)$}, represents bounded linear
combination of non-linear {components -- stellar} template spectrum {$T(x)$
convolved with a line-of-sight
velocity broadening
function, an AGN continuum model {$C(x)$}, a sum of narrow {$N(x)$} and semi-broad
components {$B(x)$} of [\ion{O}{3}]} emission lines, respectively, a sum of
Gaussian/Gauss-Hermit functions
{$\S(x)$}, accounting for other AGN emission lines in analyzed spectral domain, and
{\ion{Fe}{2}} template consisting of four groups of {\ion{Fe}{2}} lines. 

A multiplicative polynomial {$P(x)$}, that  represents
a linear combination of
Legendre polynomials, was { included} in a fit, in order to {eliminate} overall shape
differences between the observed stellar and galactic
spectra. The introduction of this polynomial in the fit
ensures that results are
insensitive to the normalization, Galactic extinction and the flux
calibration of a galaxy and stellar
template spectra \citep{Kol08}.  
For simplicity, we assumed a Gaussian velocity broadening function
{$G(x)$}, but it is possible to
use also Gauss-Hermite polynomials \citep{RW92,vanM94}. 
The contribution of the components to the total flux can be {obtained}
from their weights
which are determined at each Levenberg-Marquardt\citep{Marquardt63} iteration, using a bounding
value
least-square method \citep{LH95}. 

\begin{figure}
\begin{center}
\textbf{(a)}\hspace{.5cm}%
{\includegraphics[width=.99\columnwidth]{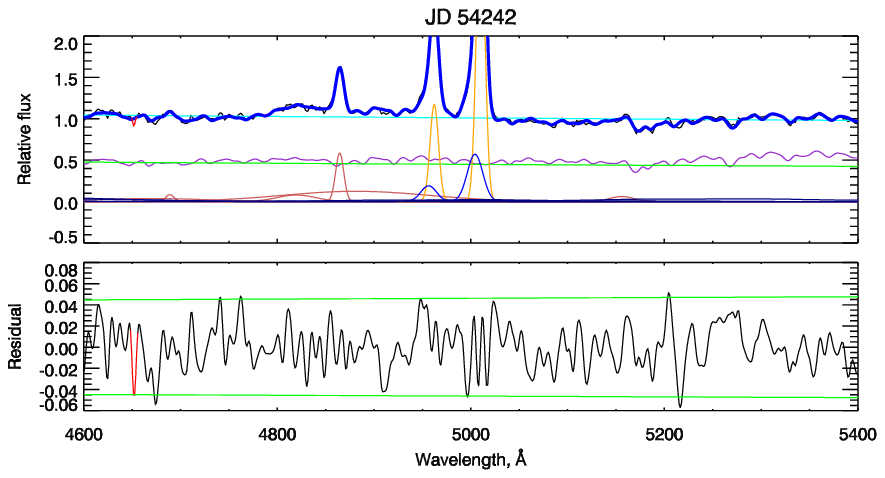}}
\\
\textbf{(b)}\hspace{.5cm}%
{\includegraphics[width=.99\columnwidth]{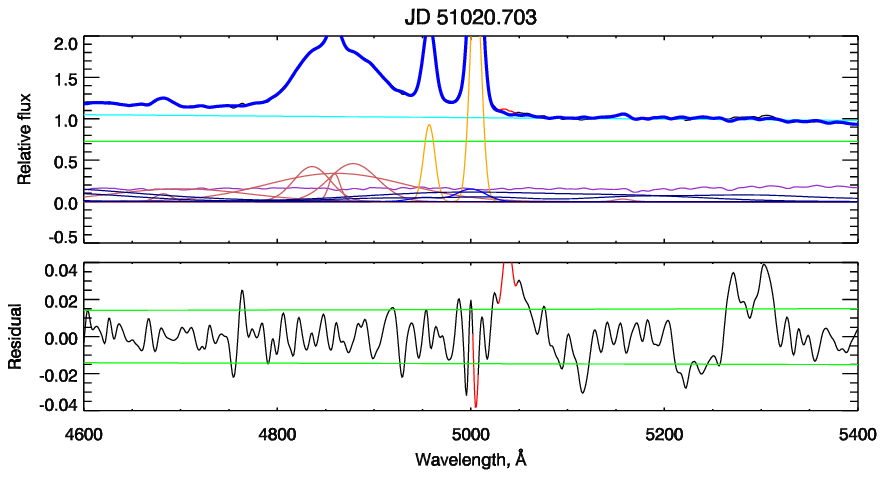}}
\caption{Examples of best fit spectra at minimum (a)
and maximum (b) {activity}. The black line in the
upper {panels of both (a) and (b) represents} the input spectrum,
the blue line represents the {best-fit} model, and the cyan line represents {the} multiplicative
polynomial, while red, yellow, blue, dark blue and violet lines represent components of
the {best-fit} model: violet -- stellar population, red-- components of {\ion{He}{2}} and H$\beta$
lines, yellow -- narrow {[\ion{O}{3}]} lines, blue -- broad {[\ion{O}{3}]} lines, dark blue
-- {\ion{Fe}{2}} multiplets and green -- {the} AGN
continuum. The bottom panel {shows} residuals from the best fit (black line).
The green solid line shows the level of the noise.
}\label{Fit_nat}
\end{center}
\end{figure}

For the stellar population model we used grid of PEGASE.HR {single stellar populations},
computed with the Elodie.3.1 library and a Salpeter IMF
\citep{LeBorgne04}.  
The model {$M(x)$} is generated at the
same resolution and with the same sampling as the observation and
the fit is performed in the pixel space. The fitting procedure performs
the Levenberg-Marquardt minimization \citep{Marquardt63}.
In modeling the integrated spectra of NGC 5548 we added
the powerlaw to the stellar
population base to represent an AGN
featureless continuum {($f_{\lambda} \ \sim  \lambda^{\alpha}$)}.
The spectral {index $\alpha$,} depends on the continuum slope, and
represents the free parameter in the fit.  
In order to tie the parameters of {the }[\ion{O}{3}] lines, we defined two 
separate components of the {model: a narrow component and a semi-broad component}. In this way we
tied the widths, shifts and intensities of {[\ion{O}{3}]}
 components (intensity ratio was kept to 3:1
).
The rest of emission lines in the domain $\lambda\lambda[4430,6400]$ were
fitted with a sum of Gaussians - we fitted {\ion{He}{2}} 4686 \AA\ with two components,
while we used four components in the fit of H$\beta$ line -narrow, broad
blushifted, broad redshifted and very broad component.
As proposed by \cite{Kovacevic2010}, we defined four groups of {\ion{Fe}{2}}
lines to fit the Fe II multiplets around the H$\beta$ line - {\ion{Fe}{2}}
s, p and f
group and I Zw1 template.
Since it is very difficult to fit stellar population in the spectra
of Sy1 galaxies with broad emission lines, in the first step we fitted
spectra of NGC 5548
in the minimum of activity with all free parameters of the model
(defined with equation \ref{eq:1}), and with multiplicative
polynomials of the 15th order \citep[as in][]{Bon2014}. We find that { a single stellar population}
with age of
7200 Myr and metallicity [Fe/H]=0.2 fits the best spectra in the minimum.
In fitting the rest of the spectra from monitoring campaigns of this
galaxy, we fixed age and metallicity at best fitted values from the first
step, and used the
multiplicative polynomial of the first order, to minimize the effects of this
polynomial on the fit of emission lines. Free parameters
associated with the stellar population were kinematic parameters - mean
stellar velocity and dispersion.
Examples of best fit spectra at the minimum and maximum
activity are presented in Fig.\@ \ref{Fit_nat}.

\begin{figure}
\begin{center}
\includegraphics[width=.99\columnwidth]{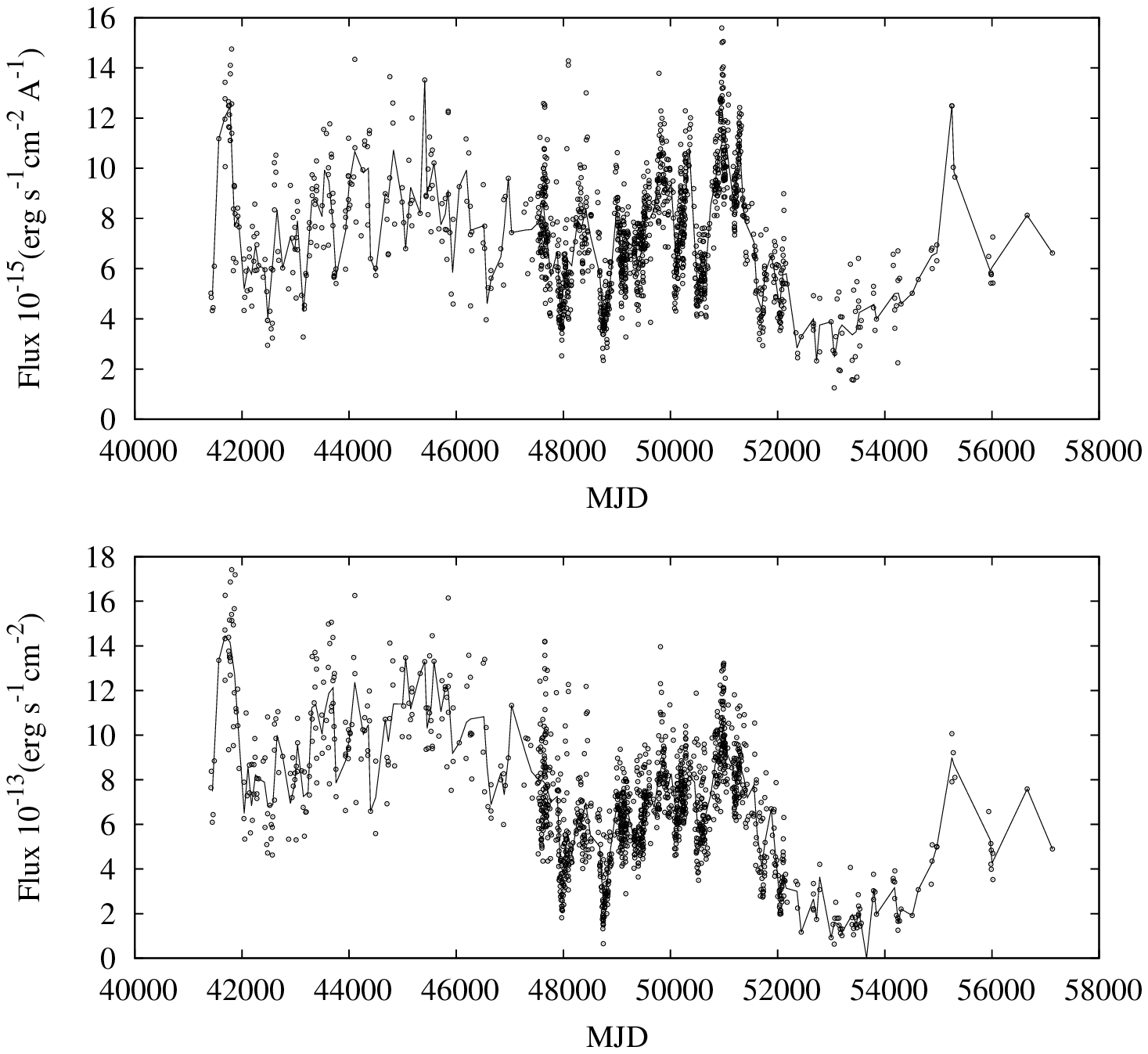}
\centering
\caption{Top: Continuum 5100 \AA\  light curve (dots) and
80 day binned continuum light curve (slashed line).
Bottom: The same as in the top panel, but for
H$\beta$.}
\label{lc}
\end{center}
\end{figure}

All the results presented below are based on the assumption that the luminosity
of the {[\ion{O}{3}]} 5007 line, and all other narrow emission lines, do not change in
time \citep[see for example][]{Peterson02,shap08}.
We assumed that the slit width and position angles used in the various  campaigns were
not affected with the {\ion{O}{3}} region size in the NGC 5548, which appeared to be
very compact \citep[see][]{Schmitt03}.
 For that reason we made a simple test to estimate light loss for smallest slit
used (see the appendix).
The spectra were scaled to a constant flux of
 F({[\ion{O}{3}]}$\lambda$5007) = 5.58 $\cdot$ 10$^{-13}$ ergs s$^{-1}$ cm$^{-2}$,
and we constructed
host galaxy subtracted continuum light curves of continuum flux measured at 5100\AA\
and broad H$\beta$ emission line.
The 5100\AA\ continuum light curve and the H$\beta$ light curve
are presented in Fig.\@ \ref{lc}.

From the set of 1494 {``IAW"} spectra,
we remove 128 spectra of the H$\alpha$ region only, 109 {spectra} with poor spectral resolution and 48 spectra which did not cover large
enough range
to measure the 5100\AA\ continuum.
This left us with 1209
{IAW} spectra, that combined with 249 spectra from Crimean
Astrophysical observatory, 83 spectra from SAO/GHO observatories, and 7 spectra from Asiago, make in total 1548 spectra. After
careful inspection of the fits, we have removed ten additional spectra,
leaving 1538 spectra for further analysis (Table\@ \ref{TAB}).

The zero point error in the radial velocity has been measured from the scatter of the difference between the 
{peak, $v_r$\,}
 of [\ion{O}{3}] $\lambda$5007 and
the narrow component of H$\beta$, $\delta v_r = v_r (H\beta) - v_r(OIII) \approx 35$ km/s.
We note that there is a known small systematic offset of
$\delta v_r \approx +19$ km/s, frequently found in {[\ion{O}{3}]}-strong AGN \citep[{e.g.,}][and references
therein]{Eracleous2003, Hu2008, Komossa2008, Marziani2016}.
The { velocity} zero point was set on the peak wavelength of {[\ion{O}{3}] $\lambda$}5007,
as {[\ion{O}{3}]  $\lambda$}5007 is a very sharp feature of high S/N, less influenced by the
underlying broad H$\beta$ profile.

Typical errors on  light and radial velocity curves have been
estimated from the dispersion of the measured parameters on short time
intervals ({$< 20$} days), in cases multiple spectra are available
(5 for the continuum light curve).
Since 20 d is a period long enough to {include possible significant} intrinsic
variations, we considered the time behavior of the {\em minimum} dispersion
value  around the average of a given parameter ({for example, }FWHM, HWs)
computed over 20 d as a function of time.
Typical rms errors for H$\beta$ seem to be around $5 \cdot 10^{-13}$ {erg s}$^{-1}$ {cm}$^{-2}$,
while for the typical rms scatter for the 5100\AA\ continuum flux is
$10^{-16}$ {erg s}$^{-1}$ {cm}$^{-2} A$.

Recently, {a paper by \cite{Li2016}} discusses a {large} part of {a} similar data set in a somewhat similar way.
The main differences between {\cite{Li2016} and our analysis is} in {our unified approach and in the quantity of } data used. In {
our study} this {was achieved} using {a}
more robust method {and} taking into account some important components that {were not considered by} \cite{Li2016} (the galactic
host emission, {\ion{Fe}{2} multiplets} and He emission and {two} components of each {[\ion{O}{3}]} narrow line in {spectral
fit}). As mentioned {above}, 1538 spectra were
analyzed (with the {new spectra} covering the
{last} 12 years), instead of only about 850 used {by} \cite{Li2016}, with a large gaps
in {the} last 12 years of their {time series}.

\section {Results}\label{results}

\subsection{Variability Analysis}\label{lc-analysis}

Light and radial velocity curves were analyzed for possible periodicity
using standard methods, such as Lomb-Scargle \citep{lomb76,scar82},
and also with the new
method for unevenly sampled data tailored to this specific case,
with the specific special conditions of our data series.
Besides light curves, we constructed curves using measurements of
{different
fractional intensities of blue side (blue dots) 
and red side (red dots) H$\beta$ broad emission line at
25\%, 50\%, 75\% and 90\%
as a function of time. They behave as half widths radial velocity curves at these fractional intensities.
We calculated line centroids\footnote{Centroids are calculated {as} $\lambda{_c}=(\lambda{_{red}}+\lambda{_{blue}})/2$}, and full
widths\footnote{Full widths are calculated {as} $\lambda{_{FW}}=(\lambda{_{red}}-\lambda{_{blue}})$} as well. Half widths and centroid
radial velocity curves are presented in Fig.\@ \ref{fig2590}.}
Full widths of the line are also calculated and their curves are presented in the Fig.\@ \ref{FW}.
We applied different methods to test for possible periodicities,
similarly as for light curves.

\subsection{Lomb-Scargle periodicity analysis}

Using Lomb-Scargle \citep{lomb76,scar82} analysis (LS method),
we analyzed light and radial velocity curves,
with previously removed
linear trends.
Results could be seen in
Table\@ \ref{LStbl}.

\begin{figure}
\centering
\includegraphics[width=.99\columnwidth]{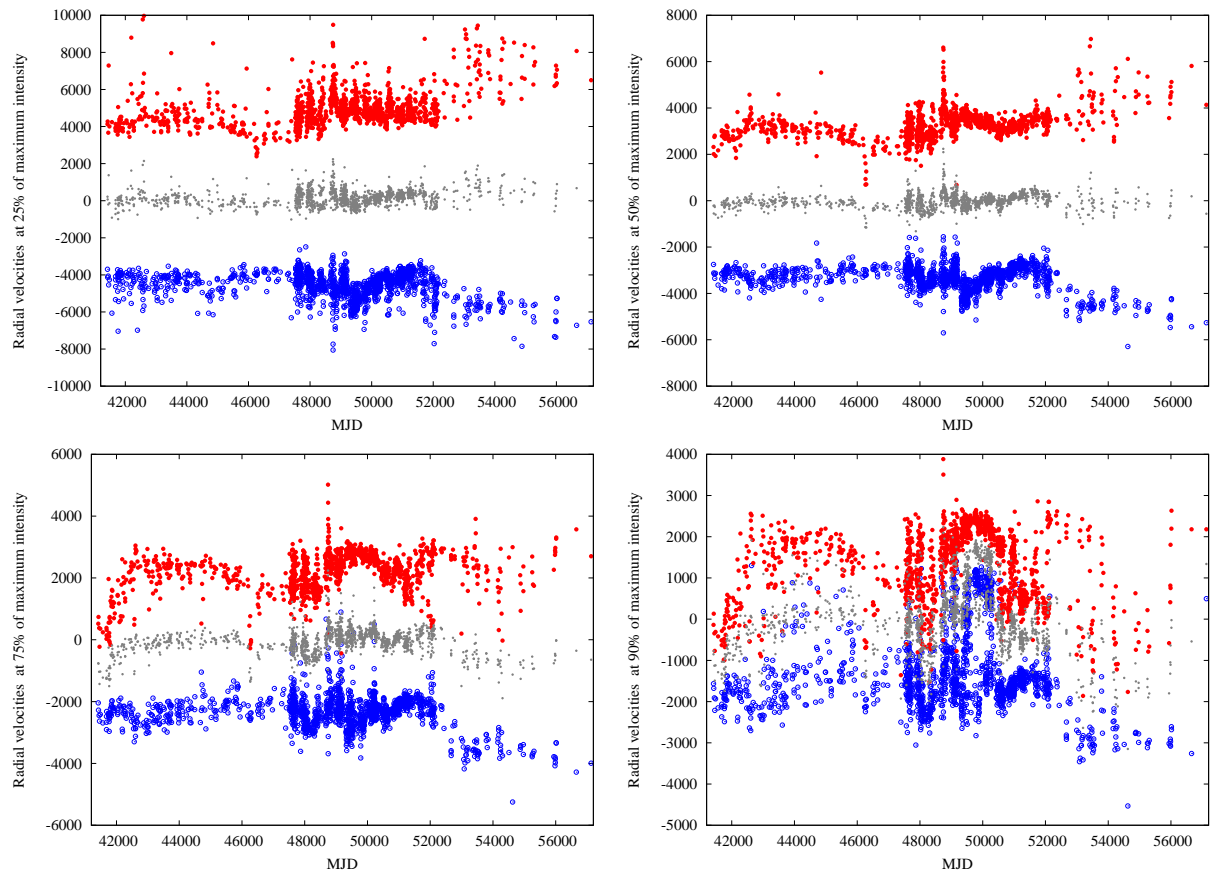}
\centering
\caption{
Radial velocities measured on the blue (open circles) and red (filled circles) sides of H $\beta _{\rm BC}$ at fractional intensities of
25\% (top left), 50\% (top right), 75\% (bottom left) and 90\% (bottom right). Centroids (gray dots) are
calculated as averaged value of blue and red side fractional intensity radial velocity measurements.
}
\label{fig2590}
\end{figure}

To avoid the problem of very different sampling, we analyzed rebinned curves. With rebining we also eliminate shorter variations that
correspond to the 
light crossing time scale of the system.
We rebinned the radial velocity curves to the 80 day average bins, since the variability lags
of H$\beta$ to continuum variations were under
30 days \citep{Zu11, Peterson02, Kaspi2000, koshidaetal14}.
According to \cite{Czerny99} there are two clearly distinct
physical mechanisms of variability in NGC 5548 curves,
with two different timescales, one under 30 days and another
with order above hundred days. The short timescale variability
is connected to the Comptonization of the soft photons emitted
by the innermost part of the accretion disk, while
in the long timescales the optical variability is not related
to X-rays \citep[see more in][]{Czerny99}.
Therefore, we assumed that 80 day binning would be long enough
to filter short variations and to analyze only longer ones.
We find
periodicities with very low false alarm probability
in radial velocity curves of half widths measured at
25\%, 50\%, 75\%, 90\% of H$\beta$ line maximum, see Table \ref{LStbl}.
As could be seen the obtained periodicities
show similar values.
We also, searched for periodicity in radial velocity curves of the full widths at
25\%, 50\% and 75\% of maximum intensity of the broad H$\beta$ emission line.
The results are presented in Fig \ref{LSFWHM}. We can see that there is a
significant peak at about 3000 days, which is about a half of the
value detected at half widths radial velocity curves (see Table \ref{LStbl}).
{As argued below, these periodicities are far from being sinusoidal and cover a
very small number of repeating periods (a little less then  3).
We suspect that the standard LS analysis may not be reliable enough in
such cases, and searched for a more robust method that is more
suitable for the type of data discussed in this paper.
The method is explained below and the numbers listed here that are based on the standard
LS method should be regarded as "tentative periods". 
We note that the LS method is
known to give spuriously high significance levels to low frequency
periods for “red noise” variability \citep[see][]{Westman2011, Vaughan2016, Bon16}
}

\begin{figure}
\begin{center}

\includegraphics[width=.99\columnwidth]{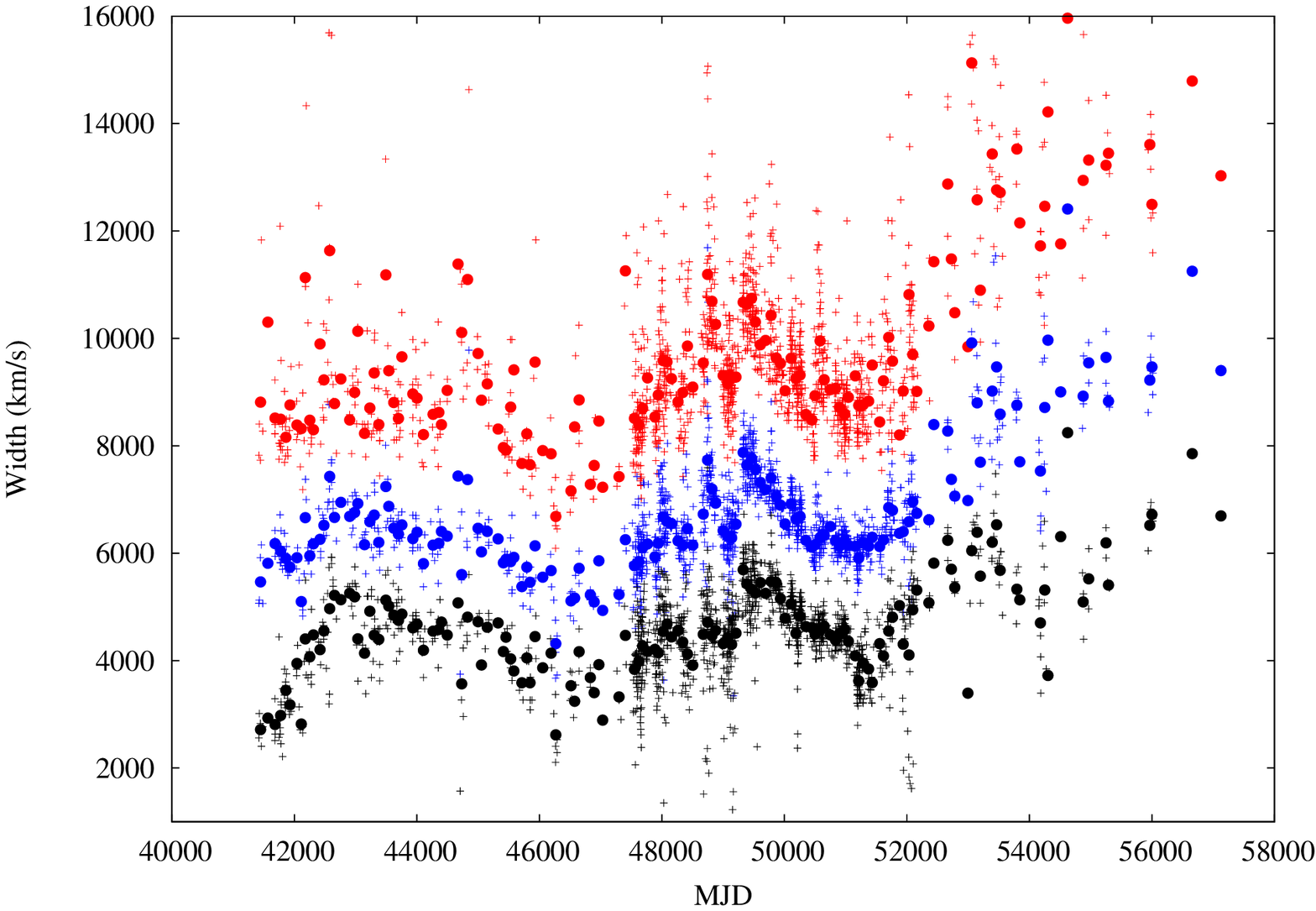}

\centering
\caption{Curves of the full width at different heights of the broad H$\beta$ line in \AA: 25\% (red),
50\% (blue), 75\% (black). The observations are presented with crosses,
while the binned averaged data are plotted with full circles.
}\label{FW}
\end{center}
\end{figure}

\begin{figure}
\begin{center}
\includegraphics[width=.59\columnwidth]{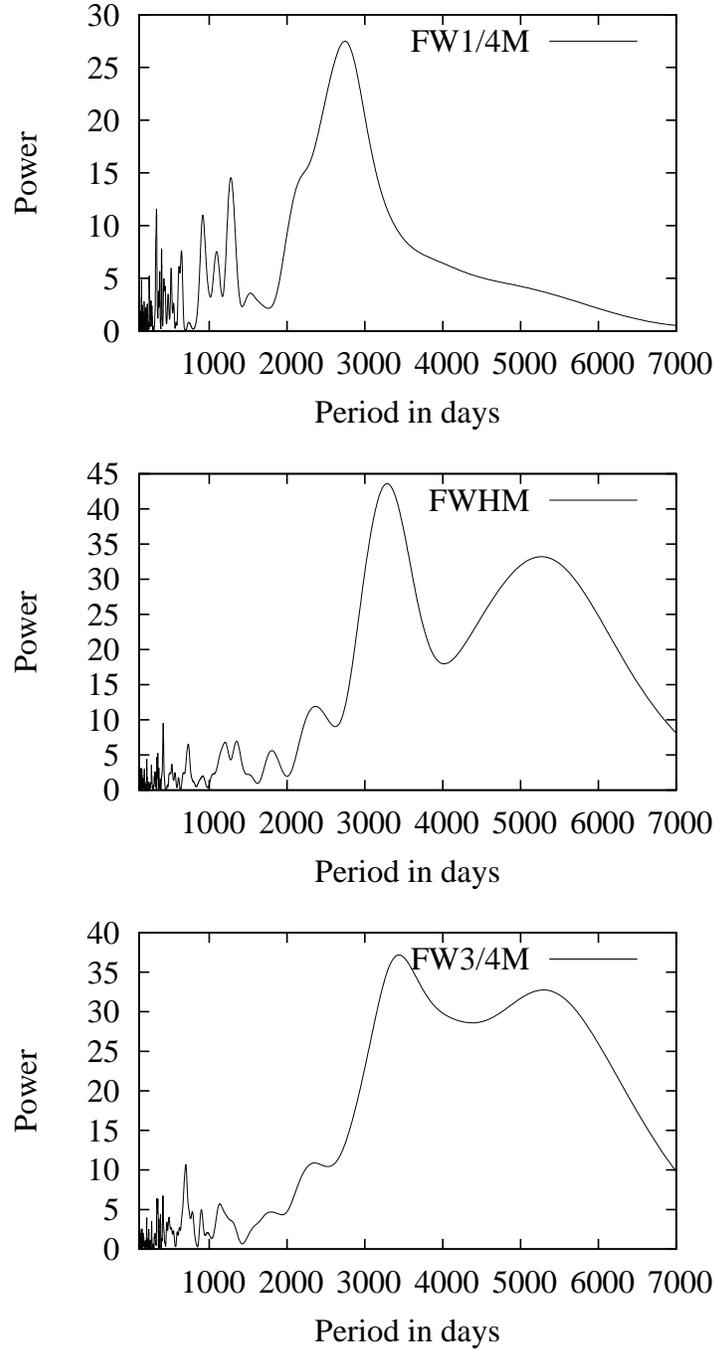}
\caption{Lomb-Scargle periodograms of radial velocity  curves of full widths
at 25\% (top panel), 50\% (middle panel), 75\% (bottom panel).
}\label{LSFWHM}
\end{center}
\end{figure}

\begin{figure}

\includegraphics[width=.69\columnwidth]{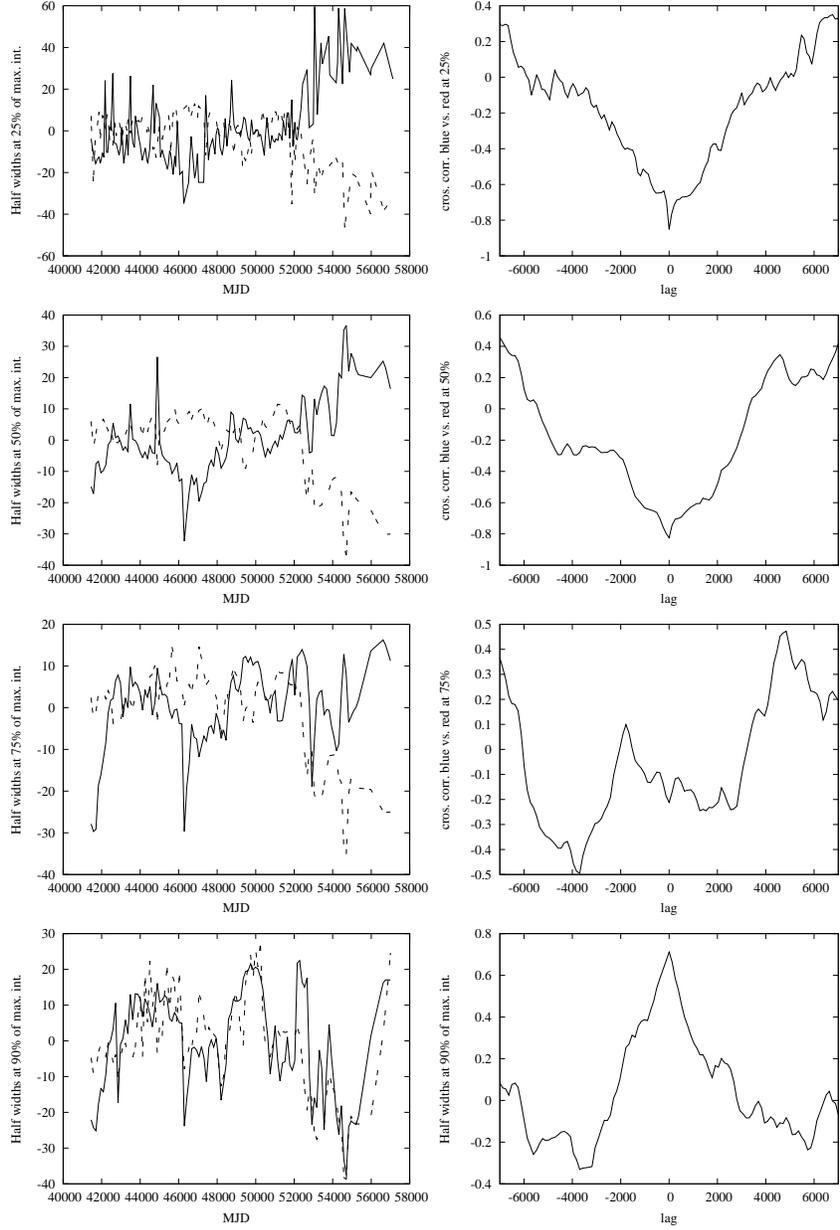}
\centering
\caption{Left: Radial velocity curves measured as red side width (thick line)
and blue side width (slashed line) of the broad H$\beta$ emission line, at
25\%, 50\%, 75\% and 90\% of the line maximum, rebinned to 80 days
with mean value subtracted, over plotted to show shape similarity. Right:
Cross-Correlation functions between corresponding pairs of curves.
The cross correlation functions of red and blue side radial velocity curves at
25\%, 50\%, 75\%  broad H$\beta$ emission line are negative (anti correlated),
while at 90\% they are correlated. This indicates that the peak of the line
measured at blue and red side on 90\% is shifting as a single component, while
at the base of the line radial velocity curves could be affected by two
oppositely moving components.}\label{figcorrel}
\end{figure}

\begin{figure}
\includegraphics[width=.99\columnwidth]{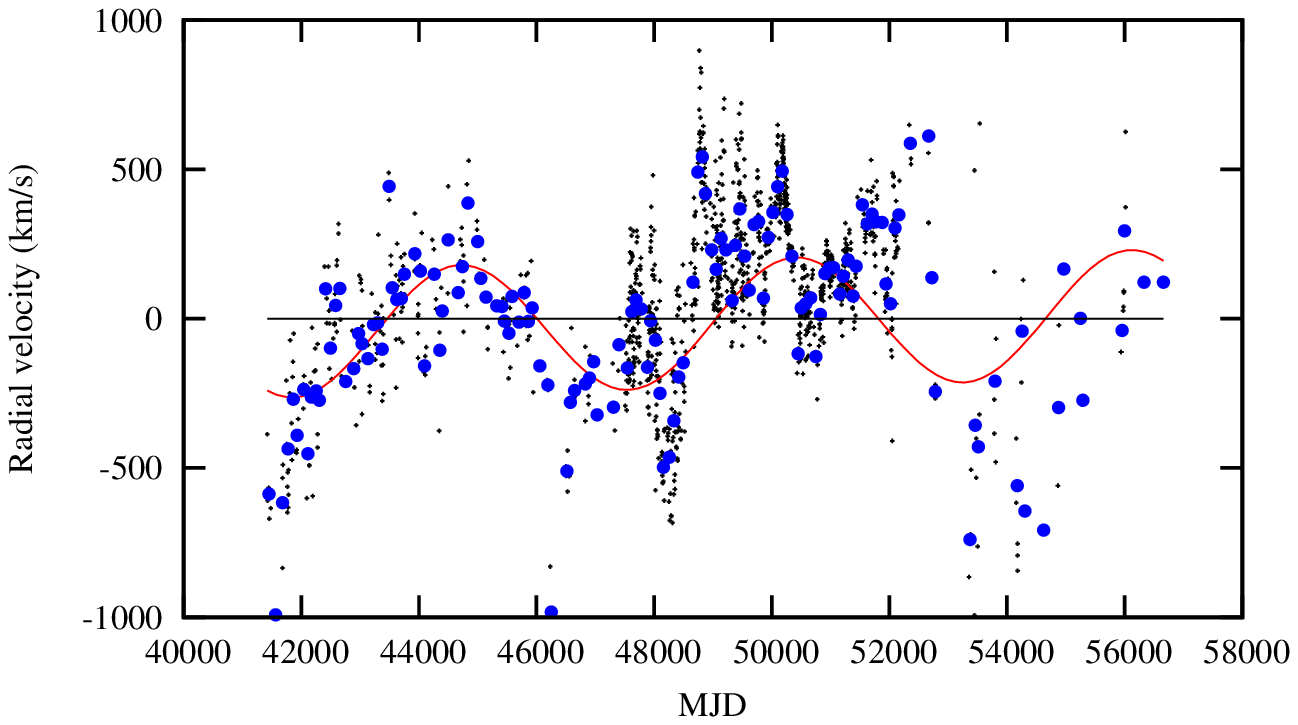}\\

\centering
\caption{Radial velocity curves resulting from
fitting a
Gaussian to the broad H$\beta$ line of NGC 5548 as discussed in the text.
The solid
red line shows the best fit of a sine wave of period 5700 days.
}\label{1G}
\end{figure}

It is interesting to note red and blue side radial velocity curves at 25\% and 50\%
are anti-correlated, while red and blue radial velocity curves at 90\% show a positive correlation,
as could be seen in Fig.\@ \ref{figcorrel}.
As we go toward the  top of the line,
the anti correlation switches to correlation, see Fig.\@ \ref{figcorrel}.
{This indicates that the peak of the line
is shifting together with the red side of the line,
while at the base of the line radial velocity curves could be affected by two
different kinematic components.}

{
To analyze the H$\beta$ line shifts we also obtained
single Gaussian fit of spectra and constructed a radial velocity curve from the
obtained shifts.
We fit a sine function, assuming the expected periodicity of 5700 days (see Fig.\@ \ref{1G}),
just to lead an eye, and not to claim a
simple sinusoidal periodicity,
since there are obvious
deviations from the sinusoidal curve in several epochs.
One can see some similarity of this radial velocity curve and the one obtained from measurements of red half width at 75\% of the line
maximum (see Figs. \ref{1G} and \ref{fig2590}), implying that the line shifts are mainly affected by variations on the red side of
the line. Also, similarity of these curves could indicate the same periodicity.
}

\subsection{A new method for finding periodicity in unevenly sampled data}\label{NEW}

\begin{figure}
\centering
\includegraphics[width=.79\columnwidth]{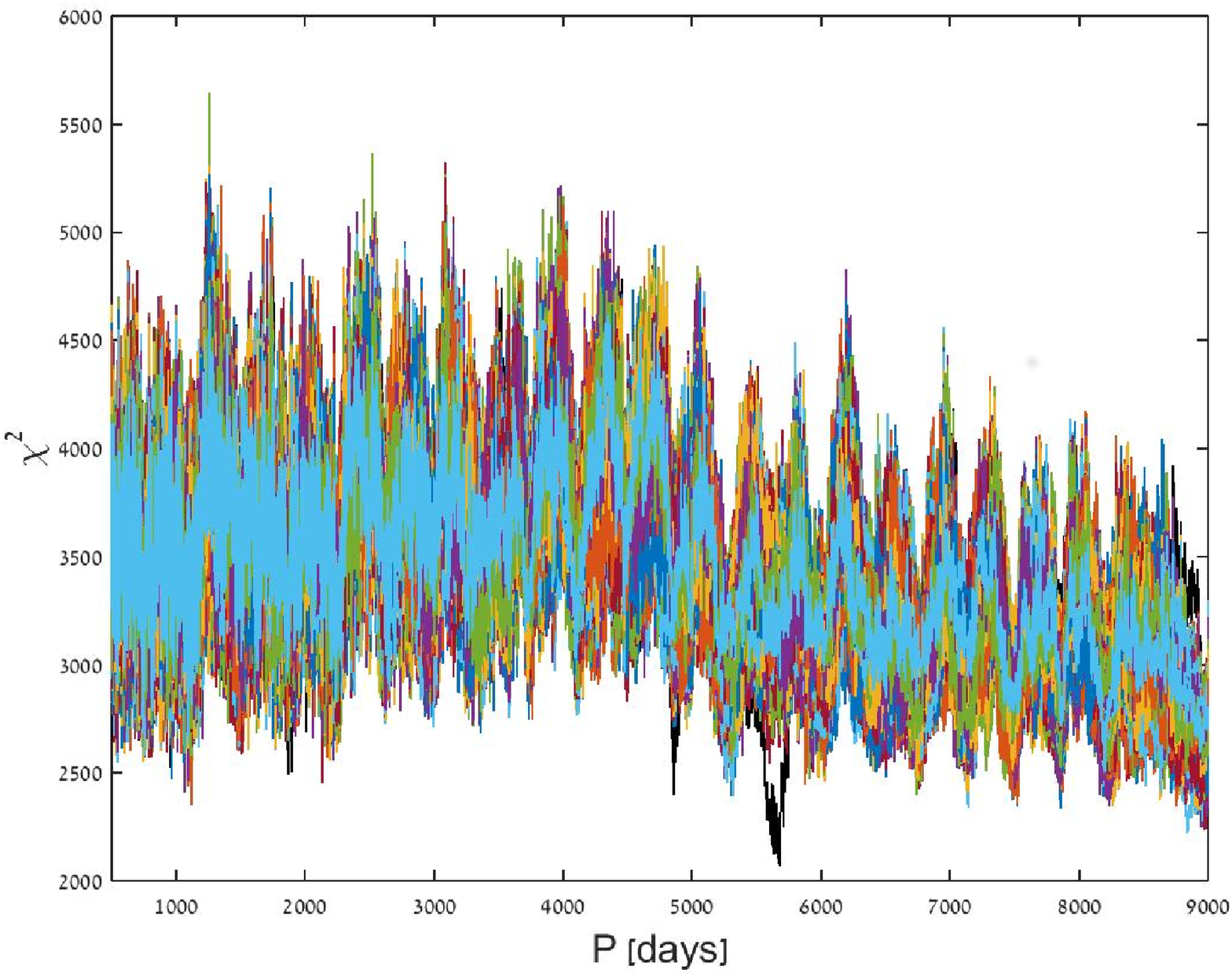}
\includegraphics[width=.79\columnwidth]{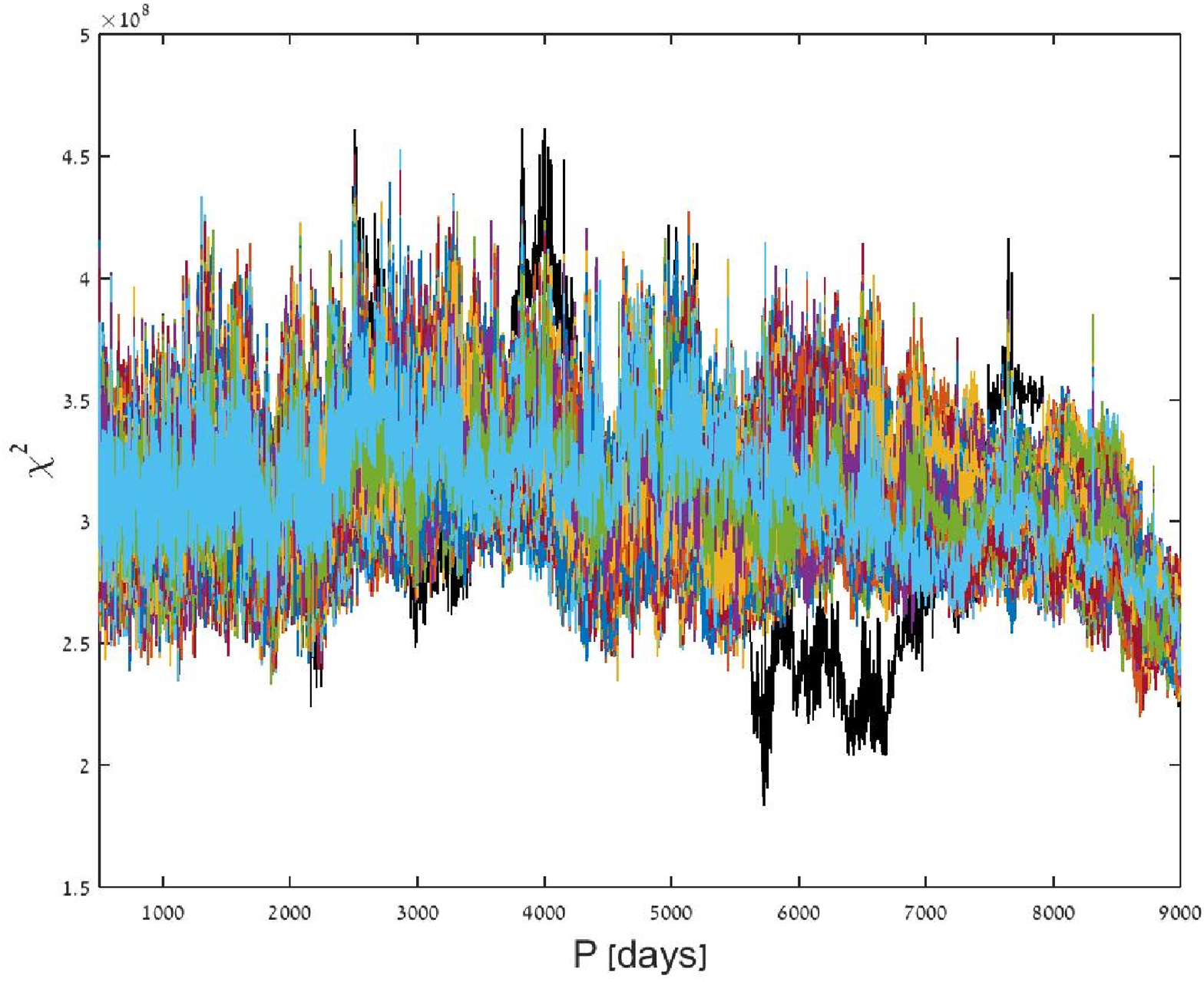}
\caption{Periodogram of the 1000 permutations
for continuum light curve at 5100\AA\ (top) and radial velocity curve measured at 75\% of the
maximum line intensity at the red side of H$\beta$ (bottom).}\label{periodogram}
\end{figure}

As explained, our data are obtained from many different monitoring campaigns
with very different sampling patterns. From the light curves one can see that
the number of data is much higher in the second third of the observed interval,
then compared with the first and third parts that contain
less than few hundred observations each, while the second third contained above
thousand spectra. Such distribution of the data introduces various biases
in standard methods for periodicity analysis of
unevenly sampled data based on sine function decomposition as
Fourier and Lomb - Scargle. 
{The Lomb-Scargle method is commonly used to detect periodicities in unevenly
sampled time series. LS follows the approach of Fourier methods and fits a
sinusoid to the data. However, in classic Fourier methods, uniform sampling gives
rise to orthogonality properties of the trigonometric functions, which have profound
statistical implications. Thus, the Fourier series terms can be shown to be
statistically independent under very broad conditions. This is no longer the
case in LS of unevenly sampled data. Nevertheless, it is a common practice
to interpret the LS periodogram as a kind of Fourier series, relying on the
assumption that the uneven sampling times are uniformly distributed. Sometimes
it is possible to neglect this effect if many cycles are included in the
analyzed time span. In our case it is clear that the sampling has an obvious
nonuniform nature, where the middle part of the time series is sampled much
more densely than the rest. Moreover, we focus our attention on long periods,
meaning only 2-3 cycles in total. Thus, the simplifying assumptions allowing the
use of LS are no longer valid, and we have to tailor a method to the specific
problem, that will take into account the nature of the specific sampling pattern.}
As could be seen comparing Table \ref{LStbl} and Fig.\@ \ref{LSFWHM}
even though Lomb--Scargle analysis
gives very high probability of periodicities,
the values of obtained periods are dispersed.
Periodicity result values span between 5500 and 6200 days.
Obviously, very unevenly sampled data required a method that would be able
to recognize repeating patterns, and not only sine functions (as in cases of
Fourier or Lomb--Scargle analysis).
For that purpose we proposed method tailored to this specific case,
with the special conditions of very few cycles, and specific sampling
characteristics.

We started from the working
assumption that the data between JD=47508
and JD=52175 are well sampled and of better quality, than
the rest (see {for example, }Fig.\@ \ref{lc}).
Those two problems prevent proper use of conventional period search techniques,
since they usually have a hidden
assumption of a homogeneous data quality and sampling, and many cycles.
Our new method treat the 'best sampled' part as fixed,
and tests only the noise and sparsely sampled parts on both ends, to see how
well they fit the hypothesis of a period. We first linearly interpolate the
'best sampled' part. Then we produce a periodogram by trying many periods. For
each trial period we 'fold' every point we test into the best sampled part, and
calculate its deviation from the interpolation (similar to the phase dispersion minimization method,
\cite[see][]{Stellingwerf1978, Plavchan2008}. We sum those squared
deviations and obtain a significance score, similar to the  ${\chi}^2$ score for the
periods. If there is a periodicity it should come up as a lower than usual
value of this score.

We look for a minimum (that is basically a ${\chi}^2$, but without normalization
by the errors), and  found it at a period of
5676 days (see Fig. \ref{periodogram}).
The question remains how statistically significant is this minimum.
To check it made the following test:
We consider the central segment as reference that we do not change.
and test whether the points in the first
and third segments actually fit the central part after phase - folding.
For this we carry out the following permutation test. 
In each iteration we randomly reshuffle
the values of the points in the first and third segment (the measurements,
not the sampling times). Thus we ``ruin'' their time dependency, but
keep their ``window  function''. We then calculate the periodogram again,
and look for the minimum value. The idea is to check what is the fraction
of random reshufflings that will produce a value lower than the value
we got with the actual lightcurve.
Only 2 out of 10000 values were lower. This means this results has a
significance p-value of around 0.0002.

The Fig.\@
\ref{periodogram} shows the actual periodogram in black, against the background
of 1000 periodograms out of the 10000. One can see that the structure of the
problem imposes some structure on the periodograms, but the result stays
significant, since the 5676 period the $\chi^2$ trough looks
quite displaced from its randomly produced counterparts.

We repeated the analysis on the radial velocity curves shown in Fig.\@ \ref{fig2590}
and show the results in the bottom panel of Fig.\@ \ref{periodogram}.
One can see that the periodicity looks very significant
on the radial velocity
curve measured at red side  75\% of the line maximum ,
the lowest point (i.e. best fit), is at a period of 5725 days - essentially
identical to the 5676 day period in the flux.
As judged from the new method, none of the other radial velocity curves shown
in Fig.\@ \ref{fig2590} shows a statistically significant periodicity.

It is important to emphasize that
real radial velocity periodicities may well be present in the H$\beta$ profile
but the noisy half width radial velocity curves do not allow us to detect them unambiguously.

{If we assume that the periodic component is sine-like, with periodicity of 5676 days,
the variance \cite[see eg.][]{Nandra97, Nikolajuk2004}
of the data carried by the periodic component (with a secular trend) is
about 19\%. We consider both the variance in the data and the
secular component under the assumption of long time scale (15.7 years) sinusoidal variations
}

\section{Possible interpretations}\label{interpretation}

The main results of {our} investigation are the detections of significant {
periodicities in the the luminosities of the 5100\AA
continuum and H$\beta$\ light {curve}, as well as in the radial velocity curve of the H$\beta$\ profile.
The periods are very similar and consistent with {$P \approx 5700$ days ($\approx 15.6$ years) - see Table \ref{LStbl}}.
This value is practically identical to the period  found in the
supermassive binary black hole (SMBBH) candidate NGC 4151 {\citep[of 15.8 years, see][]{Oknyanskij78, Bon12}}
and similar to 11 year  period of OJ287 \citep{Valtonen12}.
It is about twice the periodicity found for the case of another recently found supermassive binary candidate,  PG 1302-102
\citep{Graham15a}.
Our best measured periodicity is similar to the {14-year} periodicity found  recently
by \cite{Li2016} for {NGC~5548}, but our result is based on a more robust analysis and more data.
In {this section} we discuss, briefly, several possible physical scenarios
that can give rise to the observed periodicity. We focus on
the more secure results obtained with the new method of simulations
presented in section  \textbf{\S\ref{methods}}. We consider these
results as the most reliable and the ones obtained with the LS method
more questionable. This does not exclude the possibility that some of
the LS results, like the suggested periodicity of the FWHM velocity curve,
are not part of the suggested scenario.
In all our models we assume {a BH mass based on the results of via multiple
	reverberation-mapping campaigns.  We adopt}
$5.73^{+0.25}_{-0.24} \times 10^{7}$
M$_\odot$  {as given in the BH mass data base of \citet{bentzkatz15}}
and our own estimate of the normalization factor $f\approx$ 3.75\ used in expression
$M_{BH} = f \sigma_{rms}^{2} r_{ct} / G$, with the velocity
dispersion computed from the rms spectrum.  This mass differs by
{only 30\%\ from the mass given} by
\citet{Bentz07} who assumed $f =
5.5$ and  6.3 $\times 10^{7}$\ M$_{\odot}$.
A circular orbit
around such a BH would {have} a period {$P \approx 17.5
	r_{15 ld}^{\frac{3}{2}} \times M_{5.7 10^{7}}^{-\frac{1}{2}} \mathrm{ yr}$,}
where $r =15$ld is the radius in units of {15 ld.  This reference radius was chosen since
	it is typical of the many annual means of the lags of H$\beta$.}
Since reverberation time delays vary within {6 ld and 27 ld} \citep{bentzkatz15}
expected periods are between 8 and 36 years.  Obviously the reverberation mapping results {give a responsivity-weighted radius and the total
line emission from from a wider range of radii.}

The H$\beta$\ reverberation mapping
distances can be compared with the reverberation
measurements of the inner radius of the dust torus which range between 40 and
{80 ld} depending on observing season and on delay computation techniques
\citep{koshidaetal14}.  Finally, since some of the {models we discuss in this section}
involve a second BH, we refer to {the $5.7 \times 10^7$ M$_\odot$ black hole} as the ``primary BH" in
the system.


\subsection{Geodetic precession,  disk-self warping, hot spots and spiral arms}

{We can readily exclude}
two mechanisms that could give rise to a significant
periodicity in a single black hole system, as discussed in
\citet{Bon12}: geodetic precession,  and disk-self warping induced by radiation pressure \citep{Pring96}.
The first of these occurs on time scales that are in general much longer than
the observed periodicity \citep{Beg80}.
As for the latter, \cite{Graham15a} showed that warped disks around single black holes are not favored in
an AGN context.
In particular, the expected periodicity in such scenario is much longer  than  the
period  found for NGC 5548 (e.g., a black hole mass of 10$^8$ M$_\odot$ gives a
period between 10$^{2.2-6.9}$ years, \citealt{Pring96,Graham15b}).
Mechanisms that could produce  orbital periodicity are re perturbation
in a very large central disk that extend far beyond  the inner
accretion disk whose dimensions are known from recent reverberation mapping
to be of order 2 ld \citep{DeRosa15, Edelson2015, Fausnaugh2015}.
A hot spot rotating at a distance of about 15 ld, which is not observed during
a fraction of the orbit, can explain the periodic continuum
variations. Emissivity perturbation by spiral arms in a large central disk
can result in a double-peaked profile and emission line shape
variability \citep[see][]{Chakrabarti94, Lewis10, Jovanovic10}. Spiral arms
can be triggered by a close passage of massive object such as
massive star cluster or super massive black hole \citep[see][, and references within]{Lewis10}
by gravitational instabilities \citep[see e.g.,][and references therein]{Flohic08},
or by an object passing through the extended disk \citep{Chakrabarti1993}. Fragmented spiral arms can in
principle account for emission line shape variation on
relatively short time and radial velocity changes occurring on the
dynamical timescale \citep{Lewis10, Jovanovic10}.
Such sub-structures in a non-uniform central disk could cause an excess emission moving
across the line profile \citep{Lewis10, Jovanovic10, Goosmann2014}.
We do not consider such models as plausible explanations for NGC 5548 because of the huge central disk that is not observed in this source,
the hot spot that, in order to explain the periodic variation  should emit a sizable fraction of the luminosity of the small inner disk, and
because the overall emission line spectrum of NGC5548 is very similar to the ones observed in thousands of AGN of similar luminosity and
these scenarios cannot explain all the population properties.

\subsection{Tidal disruption events}
A tidal disruption event (TDE) {when a star is disrupted by a black hole creates an appearance somewhat} like an AGN for a limited duration.
The tidal radius (the distance from the black hole at which a
star is tidally disrupted)  can be written as
$r_{t} \approx 1.5 \times 10^{13} (M_{BH,7})^{\frac{1}{3}} (M_{\star,\odot})^{\beta} cm$\
\citep{hills75,Komossa15}, where $M_{BH}$\ is the mass of the black hole in units of $10^7 M_\odot$, and the {mass of the star} is in solar
{masses}.
The exponent  $\beta$ \ is $ \approx 2/3$\ or $1/6$\ depending on whether the star is a main sequence star  $M_{\star}$ has mass
$\lesssim 1 M_\odot$, or belongs to the upper main sequence with $M > 2 M_\odot$ \citep{Torres2010}. For the NGC 5548 black hole mass,
and for a  main sequence star  the tidal radius is extremely small, a few gravitational radii. A main sequence star may therefore orbiting
at 2-15 ld without suffering a TDE.
In the absence of a pre-existing accretion disk, tidal disruption causes a luminous flare with
a short rise, and a longer-lasting decline \citep[e.g.,][]{Rees90},
as observed in several cases \citep[eg.][and references therein]{Komossa99, gezarietal12}.
A partially stripped star in an orbit around the
SMBH can cause repeated accretion events each time its orbit passes near {pericentre} , and may
thus produce a semi-periodic {signal in the light curve} \citep[e.g.,][]{Hayasaki2015}.
However, in the case of  NGC 5548, this is
an unlikely explanation, if we were to interpret all its properties in {the context of a TDE}.
For instance, its optical narrow emission lines imply the presence of a classical narrow-line
region, and therefore a
much longer-lived AGN.
Similar arguments hold for the possibility of causing a semi-periodic
light curve from a TDE in a binary SMBH system \citep{liuetal09}.
{Another possibility is that} we may have a TDE in addition to the permanent accretion disk of a long-lived AGN,
and the TDE contributes extra accretion during each pericentre passage.
This would boost the accretion onto the BH in a periodic fashion but material must be added to the accretion flow very close to the event
horizon since the viscous time of accreting through the disk is very long. Perhaps such material is added to the central part which
increases, periodically, the X-ray emission from the disk corona which, in turn, illuminates the central disk thus boosting the optical
light emitted from its surface. X-ray illumination is known to be very important in NGC5548 \citep{Kaastra2014, Mehdipour2015,
	Mehdipour2016}, thus such possibilities cannot be excluded. A detailed discussion of such a scenario is beyond the scope of
the present paper.
{We note, however, that TDEs are very rare events \citep[one event per
inactive galaxy every $10^4-10^5$ years; e.g.]{Rees90, MagorrianTremaine1999}, even
though rates can be higher in AGN \citep{KarasSubr2007}, and in the presence of supermassive
binary black holes \citep[e.g.][]{Ivanov2005, Chen2011}.
In any case, chances of seeing such an event in only
one single nearby galaxy analyzed are very small, and we therefore consider this possibility as
very unlikely.}
Similarly, \cite{Landt15} concluded that a TDE is a very unlikely
explanation in NGC 5548.



\subsection{Binary black holes}
There are several scenarios involving binary BH systems in
{AGNs}. These can be divided into two {broad} groups. One {is} where
only one of the BHs has an accretion disk and a BLR associated with it.  {The other is the case} where both BHs are accreting through their
own disks. For roughly equal BH mass (the only case considered here)
the average separation of the two is of order 20 ld and hence there is
only one dusty toroidal structure around the two.
{
	In general, such systems are thought to be the end result of a galaxy merger,
	where the two BHs from the two galaxies are at the final stage of merging
	\citep[see][]{MM2001, MM05}.
	Earlier studies of NGC5548 suggest some evidence for a merger 0.6-1.0
	Gyr ago \citep[more details in][]{Tyson1998, Steenbrugge2005, Slavcheva2011}.
	Some simulations performed in order to characterize the SMBBH systems,   show
	the formation of a  circum binary disk, inside of which the two BHs are
	accreting matter forming mini accretion disks  \citep[e.g.][]{Hayasaki08,
		MacFadyen08, Bog08, Bog09, Cuadra09, SmaBon2015}. Further out, the {circum-binary disk}  cools and may form a torus.
{Blending BLRs,  have also been investigated
		\citep[e.g.,][]{Shen10, Pop2012}, or in case of a
		{high-mass-ratio} system, only one shifting BLR may be seen.}
	A second black hole can give rise to a host of phenomena that can yield
	periodic signals
	\citep[see][]{Katz97, Sillanpaa88, Bog08, Bon12, Kun2014, Graham15a, Graham15b}}.
A detailed scenario of this type has been investigated in several papers by Bogdanovi\'c and collaborators
\citep{Bog08, Bog09}.
{They simulated a {high-mass-ratio} system with nearly identical time
	interval and periodicity as found here.}
These involve disk disruption, the formation and destruction of
spiral arms in the gas between the BHs, and more. Some features of this
model are appealing, especially those corresponding {to periodic}
changes in the velocity curve of part of the gas, including cases where
only one side of the profile is affected. Unfortunately, there {have been} no
{attempts} to use the results of the dynamical simulation to calculated
the resulting emission line spectrum, line profile and time variations
in the systems. While we are not in a position to look into this in
detail, we note that the gas configuration in this model, and the gas
properties, might be very different from what is known from many year of
study of NGC 5548. Moreover, as noted earlier, the {broad-line} spectrum
of this source is very similar to the spectra of {thousands} of other
type-I AGN. We thus consider it less plausible that NGC 5548 contain such a
binary BH system. 

A binary BH system where only one of the BHs carries its own disk and
BLR (although the outskirts of the BLR must be disturbed by the
``naked" BH) is perhaps easier to explain. Obscuration in this scenario
is very inefficient. However, we note that gravitational lensing of
the luminous disk around the primary BH by the second BH can enhance
the continuum emission by a factor of order 1.11 over a period of a few
hundred {days, while the effects over BLR emission would be even smaller, by a factor of less then 1\%}. For example, if both BHs have a
mass of {$5 \times 10^7$} M$_{\odot}$ and their separation is 20 ld, the size of the Einstein ring is {$\approx$ 0.48 ld}.
This size should be compared with the size of
the disk at 5100\AA\ ($\approx$ 2 ld). Such an enhancement {is} achromatic {which gives} the immediate prediction that other continuum
wavelengths
would show an identical change of amplitude during the passage.
Such a scenario cannot explain the periodic H$\beta$ variations or its periodic
light curve. Moreover, 15 ld is well inside the BLR so dynamical
changes in the {line-emitting} gas must be considered too.

{Using a {radial-velocity} test for supermassive BBH for broad,
	double-peaked emission lines \citep{Liu16}
	assuming
	equal mass {components,} the line
	peaks should be at about 2300 \kms.
	At some epochs very small moving peaks in emission lines are identified
	corresponding such velocities \citep[see][]{Shap04}.
	{However,} it is very clear that the red and blue wings of H$\beta$ respond to
	the same continuum variability at roughly the same time, so the gas emitting
	them is approximately at the same distance from the primary accretion disk.
	Recently, \cite{Li2016} proposed a BBH scenario in NGC 5548,
	as a result of their two Gaussian decomposition model, fitted into 150 day averaged
	spectra. We tried to test these claims, using series of two Gaussian
	decomposition models with different types of constraints
	({e.g., a} constant intensity ratio with a significant width difference of each Gaussian,
	forcing them to {fit different} parts of the line,
	narrower fits {for} the core, and wider {for} the wings).
	We found that such {a} configuration
	could result {in} Gaussian components that
	switch sides and cross from blue to red and opposite.
	Unfortunately, in all cases we tested
	(assuming constant initial parameters of component shifts),
	there were additional crossings in radial velocity curves,
	that ruined the expected periodicity.
	We also note that in our {modeling with} a single Gaussian fit {to the}
	spectra (Fig. \ref{1G}), {we obtained a}
	radial velocity curve {appearing} somewhat similar to the
	one of 75\% red half width (see Figs. \ref{1G} and \ref{fig2590})
	{.  This could imply the possibility of}
	the same periodicity {and could} eventually indicate {the} presence of {high-mass-ratio} system.
	According to such analysis, we {again} find BBH hypothesis {to be} unlikely,
	except for the case of {high-mass-ratio} system, {which we can neither disprove nor support.}
}



\subsection{Obscuration by gas and dust inside and outside the BLR}
This category includes several possibilities of obscuration of
the central disk, and part of the BLR, by a moving object at a distance
corresponding to a 15.7 yr period. We consider two different
possibilities, one an object inside the BLR moving around the primary BH, and
{the other} an object which is part of an outflowing wind moving
around the polar axis of the disk. Both scenarios correspond to a situation where
the length of obscuration is a small fraction of the period,
perhaps a year or even less. Such a situation is consistent with the new scheme
presented in section \ref{results} which confirm the periodicity
but do not show whether it is sinusoidal in shape or whether it
corresponds to only one short event of dimming or enhancing the radiation of the central source.
A large dust-free cloud moving around the primary BH, inside the BLR, at a distance corresponding to a period of 15.7 yr, can cause periodic
obscuration of the central continuum source. This explanation is appealing since the required distance, about 15 ld, is exactly in the
middle of the range of the multi-year RM size of the Hb line and coincidental agreement between the two distances is unlikely. There are
various difficulties to this scenario. The obscuring material must by thick enough and
large enough to occult a large fraction of the central accretion disk which is 2
ld in radius, \citep[see e.g.][]{Fausnaugh2015}.
This dimension is larger than the typical size of BLR clouds that are of order 0.1
ld across assuming the density and column density
are $ \approx 10^{10} cm^{-3}$ and $\approx 10^{14} cm^{-2}$, respectively
\citep[e.g.][]{Netzer2013}.
A dust-free cloud must be Compton thick to block the  5100\AA\ continuum
since the ionized column of such gas is only of order $10^{22-23} cm^{-2}$.
Thus both the radial and lateral dimensions of such a cloud are
orders of magnitude larger than those considered typical of the BLR
(a collection of clumps adding up to the required dimensions is just as
difficult to explain).
Obscuration by a dusty gas cloud is easier to explain since a very
small column density, corresponding to $A_V<1$ is all that is required to
absorb much of the radiation at 5100\AA . However, 15 ld is well within
the dust sublimation radius for this source and such grains will not
survive in this environment. One can consider a very large dusty
cloud in a spiraling elliptical orbit \citep[e.g.][]{NetzerMarz10} where,
in this case, part of the dust is not sublimated because it is
shielded from the central source radiation during a big part of
the orbit. We did not explore this possibility in detail
but consider it problematic because efficient shielding of the dust from the
radiation at wavelengths longer than the Lyman or Balmer continuum
edges is hard to explain. Explaining the observed periodic variations in
L(H$\beta$), and the period in its velocity curve, is even more challenging.
Obscuration by dusty gas spiraling along the polar axis,
as part of a large scale disk wind, is an alternative explanation. Disk winds have
been considered for years as a general scenario to explain
both the BLR and the dusty torus structure around the BH
\citep[see e.g.][]{Elvis2000, Elitzur2006, Elitzur2008, Czerny2011,  Netzer2013, Arav2015, Netzer15, EltNetz2016}.
This geometry, which is sketched in {Fig.\@ \ref{torus}}
allows a period of 15.7 years at a distance much larger than
15 ld from the BH since material is rotating around the polar axis of the system, {with small opening angle of about 30
degrees \cite[as suggested earlier for this object, see eg.][]{Rokaki93, Kaastra2014} or less}.
For example, a dusty cloud at a distance of 50 ld, well
inside the dust sublimation radius, can obscure the central
source every 15.7 yr for a few hundred days. The required column density is
small, corresponding to $A_V<1$ mag, and the lateral
dimension of order the disk size, $\sim$ 2 ld (while we talk about ``a cloud" this may
well be a collection of clumps). Such a cloud can
also obscure part of the BLR although this fraction is much smaller because of the much
larger dimensions of the BLR.
Obscuration by dusty material must result in a wavelength
dependent reddening of the central continuum. In principle, the obscuration may
last only a few hundred days every 15.7 years and the available
spectroscopy is not good enough to exclude such short-term events of reddening.
Another difficulty with spiraling out material is that the
outward motion of the gas will result in a continuous change of radius and hence
periodicity .
Such a change may be small over the 43 years of observations considered here. All these
details are, again, beyond the scope of the present work.
The geometry of the torus considered here is very different
from the simple tori considered in earlier studies \citep[][and
references therein]{Netzer15}. This opens a range of possibilities
that may be related to the periodicity discussed here, in
particular the obscuration by line-of-sight dust. Large structures
in the non-uniform torus wall, scattering by the torus dust, etc, all should be
investigated in detail. 

\begin{figure}
	\begin{center}
		\includegraphics[width=.89\columnwidth]{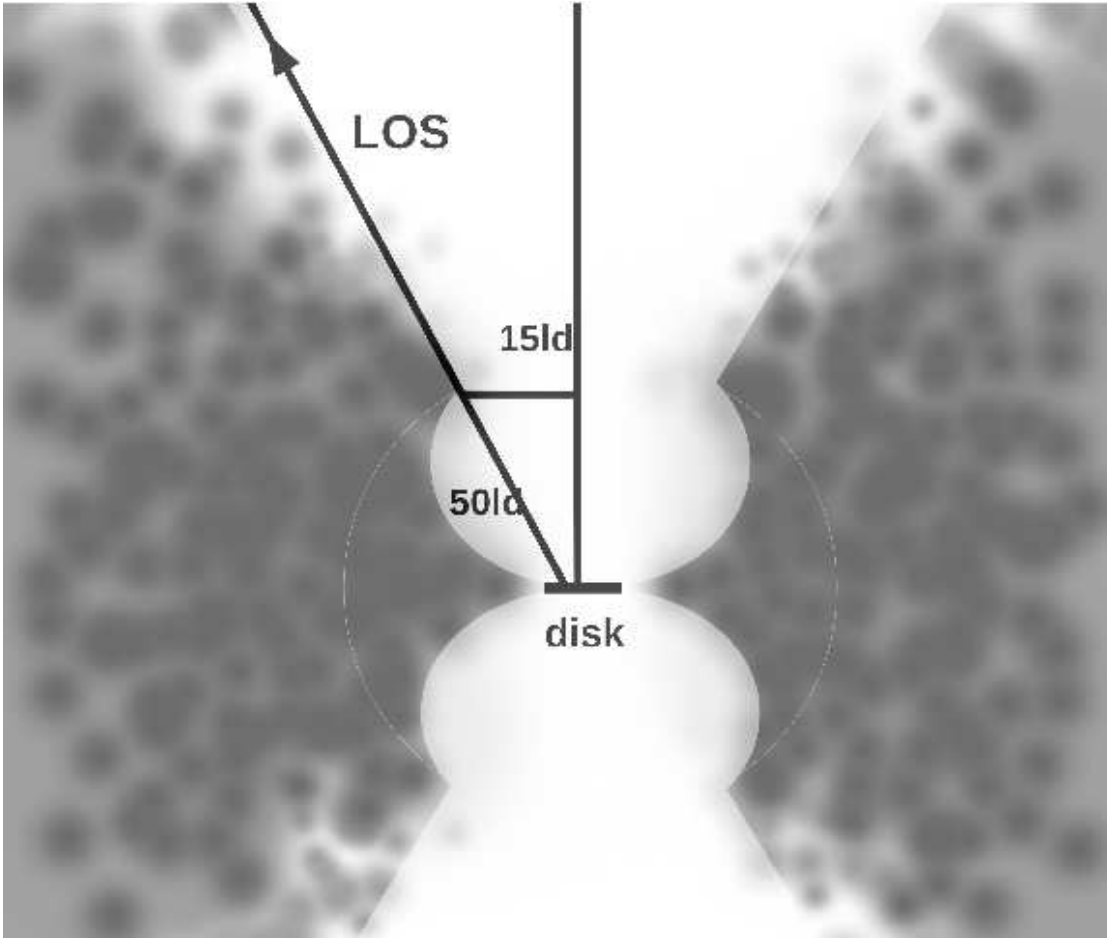}
		\centering
		\caption{A representation of the disk obscuration, where partially
			obscured disk is seen through the inner edge of the torus cone (15 ld from the
			axes and at the distance from the disk of about 50 ld). Obsucration in
			such configurations are possible if the
			line of sight (LOS) is similar to the cone opening angle. In this case it
			would correspond to angles of about 30$^{\circ}$, or less. The obscuration
			model assumed here has been inspired by various torus simulations obtained with SKIRT code
			\citep{Stalevski16}.
		}\label{torus}
	\end{center}
\end{figure}


\subsection{Periodic X-ray enhancement and reflection}
The main difficulty in all previous scenario is the lack of
explanation of the clear connection between the periodic continuum variations,
the periodic H$\beta$ flux variations, and the periodic H$\beta$ velocity variations.
A more logical explanation is a mechanism that causes 
a periodic
enhancement of
the ionizing continuum, which in turn causes a periodic H$\beta$ intensity
variations (reverberation), which will cause {time-dependent} changes
in the mean emissivity radius of the H$\beta$ line. In a virialized cloud
system, an increase in the H$\beta$ emission region is correlated with smaller
gas velocities which is in agreement with the observed line width variations.
About 2/3 of the bolometric luminosity in NGC 5548 is due to {far-UV and soft}
X-ray emission{\citep{Gaskell2008}}. The X-ray source is probably illuminating the central disk
causing much of the optical-UV continuum variations \citep[see][]{Fausnaugh2015,
	Uttley2003, Edelson2015}. 
Accretion events close to the primary BH, that
enhance the emitted X-ray flux in a periodic fashion, can trigger the
entire chain of events from optical continuum variations,
to H$\beta$ flux variations, to H$\beta$ velocity variations. As suggested above,
a {pericenter} {passage of a}{partially} tidally disrupted star is perhaps one possibility for such {an event}.
Also, {an orbiting G2 like object
\citep[see][]{Witzel2014} or some other stellar size object with an orbit in a collision with the disk are perhaps
	some possibilities that could lead to such events.
	The energetics of these kind of
	pericentric encounters depend very much on the consistency of the medium  (the
	accretion disk or the torus)  that is being crossed with their orbit.}
One can also imagine one or more gas clouds from the BLR in
very eccentric orbit the origin of such events.
However, the mass of a single BLR cloud is tiny which makes it very unstable against tidal forces and its mass supply to the BH {is many}
orders of
magnitude below what is required to produce a significant X-ray flare.



\subsection{Orbiting body crossing the accretion disk}
Periodic variations could be caused by an orbiting object perturbing the accretion disk while passing through it \citep[eg.][]{Syer1991,
	Chakrabarti1993, Armitage1996, Subr1999, Bogdanovic2016, Pihajoki2016, BogNguyen2016}.
Such scenario was proposed for  NGC 5548 in \cite{Shap04} and was
connected to the appearing  and shifting  bumpy features in the red wing
of H$\beta$. If we assume a {stellar-mass} object passing through the disk at {a} radius
smaller than 15 ld, it could cause a perturbation in the disk,
producing shocks \citep[eg.][]{Chakrabarti1993}.

{
	Such a collision could heat the disk
	\citep[see e.g.][raising the temperature
	above $10^7$ K]{Bogdanovic2016}
	producing periodic optical and X-ray emission.
	It is not our intention to attempt such calculations {for} NGC 5548, only to
	mention that the disk dimension,
	and the orbit eccentricity, are likely to be the limiting factors in such cases.
	The {hot spots caused by impacts} may not necessary need to live for the complete orbital
	time, but would be made periodically with each collision.
	Knowing that in our Galaxy {there are a} number of
	central {stars (the ``S0 stars") that} are on highly eccentric, randomly
	inclined orbits \citep{Eckart97, Ghez98, Gillessen09},
	and some of them show {} similar periodicities as here ({e.g.,} for S0-2
	{the} periodicity is 15.2 years, for S0-14 {it} is 38 years \citep[eg.][]{Zucker2006}
	({e.g., }Zucker et al. 2006), {and} for S0-102 it is 11.5 years \citep[see][]{Meyer2012}),
	it is conceivable {that one could} find objects on inclined orbits that could cross the {accretion disk of NGC 5548}.
	In {the case of a star} passing through the {part of the disk responsible for the optical emission},
	at about 2 ld radius, the eccentricity of the orbit
	would be about 0.7.
}



\section{{Conclusions}}\label{conclusions}

We have analyzed {the 5100 \AA\ continuum} and the  H$\beta$\ light and radial velocity curves of NGC 5548  using about 1600 spectra
spanning 43 years, including 12 years of
new data.
The main results of the study are as follows:
\begin{enumerate}
	\item The continuum light curve shows a of periodicity
	of about 5700 days {at a high confidence level}.
	Similar periodicities are found in the light and radial velocity curves of
	the broad H$\beta$ emission line.  
	The period has been detected through a standard periodogram analysis
	which we consider not very significant, as
	well as through a new method specifically devised for the present data set that
	takes into account its heterogeneous quality and uneven sampling.
	\item The detected periodicity is consistent with orbital motion inside the BLR of the source.
	\item We examined various physical scenarios that can explain the observed periodicity.
	These include binary {BHs, a TDE of} a massive star,
	orbiting dust-free and dusty clouds around the central BH and the polar axis of the system
	(in a polar {wind),} and periodic enhancement of the
	inner part of the disk producing the X-ray emission.
	While none of these can explain all the observations,
the preferred explanation is the one linking the enhanced {continuum, the enhanced
line emission} and the lowering of the velocity through a single scenario related to the X-ray emission in this source.
	{The enhanced X-ray emission could be triggered by an
		orbiting object periodically colliding with the accretion disk.}
\end{enumerate}


 \acknowledgements

{{
We would like to thank the International AGN Watch group for spectra 
available on their website, especially Bradley Peterson for with kind permission to use unpublished
spectra observed at Asiago.
Also, we would like to thank Tamara Bogdanovi\'c, Jack Sulentic and Mike Eraclous
for all comments and suggestions.}
This research is part of the projects 176003 ''Gravitation
and the large scale structure of the Universe'' and 176001 ''Astrophysical
spectroscopy of extragalactic objects'' supported by the Ministry of Education
and Science of the Republic of Serbia.
This work was supported by: CONACyT research grant 151494 (México), INTAS (grant N96-0328),
RFBR (grants N97-02-17625, N00-02-16272, N03-02-17123, 06-02-16843,
N09-02-01136, 12-02-00857a, 12-02-01237a, N15-02-02101) and
MS acknowledges support by FONDECYT through grant No.\ 3140518.
}

\LTcapwidth=\textwidth
\begin{tiny}
\begin{longtable}{|l|l|l|l|l|l|l|}
\caption{ \footnotesize Log of the new observations 2003-2015 from SAO,
INOAE and Asiago.
{Columns in order are}:
number of the observation; date of the observation
(DD.MM.YYYY); Julian date; a code assigned to the observatory (N); { the aperture used; the spectral region covered and the
seeing.}}
\label{Tab_SAO} \\
\hline 
\multicolumn{1}{|c|}{\textbf{  No}} &
\multicolumn{1}{c|}{\textbf{ Date}} &
\multicolumn{1}{c|}{\textbf{ JD-2400000}} &
\multicolumn{1}{c|}{\textbf{ Observatory}} & 
\multicolumn{1}{c|}{\textbf{ Aperture}} &  
\multicolumn{1}{c|}{\textbf{ Sp. domain}} &  
\multicolumn{1}{c|}{\textbf{ Seeing}} \\
\hline 
\endfirsthead
\multicolumn{7}{c}%
{{ \bfseries \tablename\ \thetable{} -- continued from previous page}} \\
\hline \multicolumn{1}{|c|}{\textbf{  No}} &
\multicolumn{1}{c|}{\textbf{ Date}} &
\multicolumn{1}{c|}{\textbf{ JD-2400000}} &
\multicolumn{1}{c|}{\textbf{ Observatory}} & 
\multicolumn{1}{c|}{\textbf{ Aperture}} &  
\multicolumn{1}{c|}{\textbf{ Sp. domain}} &  
\multicolumn{1}{c|}{\textbf{ Seeing}} \\
\hline
\endhead
\hline \multicolumn{7}{|l|}{{Observatory codes:}}\\
\multicolumn{7}{|l|}{{ \hspace{0.07\linewidth}  SAO - Special Astrophysical
Observatory of the Russian Academy of Science (Russia) }}\\
\multicolumn{7}{|l|}{{ \hspace{0.07\linewidth}  GHO - INOAE Guillermo Haro
Observatory (Mexico)   }}\\
\multicolumn{7}{|l|}{{ \hspace{0.07\linewidth}  ASG - Asiago Astrophysical
Observatory (Italy)   }}\\
\hline \multicolumn{7}{|r|}{{ Continued on next page}} \\ \hline
\endfoot
\hline 
\endlastfoot
1 & 26.01.2003 & 52665.924 & GHO & 2.5x6.0 & 5700-7360 & 3.5 \\
2 & 27.01.2003 & 52666.926 & GHO & 2.5x6.0 & 3800-7090 & 1.5 \\
3 & 28.01.2003 & 52667.901 & GHO & 2.5x6.0 & 3800-7090 & 2.5 \\
4 & 27.03.2003 & 52725.893 & GHO & 2.5x6.0 & 5700-7360 & 2.3 \\
5 & 12.04.2003 & 52741.582 & SAO & 4.0x19.8 & 5590-7300 & 2.0 \\
6 & 11.05.2003 & 52771.272 & SAO & 2.0x6.0 & 5738-8097 & 1.2 \\
7 & 23.05.2003 & 52782.728 & GHO & 2.5x6.0 & 3700-7170 & 3.6 \\
8 & 23.06.2003 & 52813.778 & GHO & 2.5x6.0 & 5600-7300 & 2.7 \\
\end{longtable}
\end{tiny}

\LTcapwidth=\textwidth
\onecolumn
\begin{center}
\begin{tiny}
\begin{longtable}{|l|l|l|l|l|l|l|l|l|l|l|l|l|}
\caption{Spectral parameters obtained from the best fits. Table lists: Julian
Date (N1); H$\beta$ flux; Continuum flux measured at 5100 \AA in units of $10^{−15} erg/s/cm^2$ and broad H$\beta$ emission line in units of
$10^{−13} erg/s/cm^2/$\AA;
and wavelengths measured at
blue and red side of 25\%, 50\%, 75\% and 90\% of maximum intensity of broad H$\beta$ emission line in units of $km s^{-1}$. Here we present
only example of the table (with beginning and the end part). 
The complete table in electronic version is appended.}
\label{Results} \\
\hline \multicolumn{1}{|c|}{\textbf{JD-2400000}} &
\multicolumn{1}{c|}{\textbf{Flux Hb}} &
\multicolumn{1}{c|}{\textbf{Err.}} &
\multicolumn{1}{c|}{\textbf{Fl. 5100}} &
\multicolumn{1}{c|}{\textbf{Err.}} &
\multicolumn{1}{c|}{\textbf{B25\%}} & 
\multicolumn{1}{c|}{\textbf{R25\%}} &  
\multicolumn{1}{c|}{\textbf{B50\%}} &  
\multicolumn{1}{c|}{\textbf{R50\%}} &
\multicolumn{1}{c|}{\textbf{B75\%}} &
\multicolumn{1}{c|}{\textbf{R75\%}} &  
\multicolumn{1}{c|}{\textbf{B90\%}} &  
\multicolumn{1}{c|}{\textbf{R90\%}} \\ 
\hline 
\endfirsthead
\multicolumn{13}{c}%
{{\bfseries \tablename\ \thetable{} -- continued from previous page}} \\
\hline \multicolumn{1}{|c|}{\textbf{JD-2400000}} &
\multicolumn{1}{c|}{\textbf{Flux Hb}} &
\multicolumn{1}{c|}{\textbf{Err.}} &
\multicolumn{1}{c|}{\textbf{Fl. 5100}} &
\multicolumn{1}{c|}{\textbf{Err.}} &
\multicolumn{1}{c|}{\textbf{B25\%}} & 
\multicolumn{1}{c|}{\textbf{R25\%}} &  
\multicolumn{1}{c|}{\textbf{B50\%}} &  
\multicolumn{1}{c|}{\textbf{R50\%}} &
\multicolumn{1}{c|}{\textbf{B75\%}} &
\multicolumn{1}{c|}{\textbf{R75\%}} &  
\multicolumn{1}{c|}{\textbf{B90\%}} &  
\multicolumn{1}{c|}{\textbf{R90\%}}  \\ 
\hline
\endhead
\hline \multicolumn{13}{|l|}{{Note: (1) Julian dates of IAW watch data in the
form ``*.1" are used to artificially make a difference between two
different
spectra under the }}\\
\multicolumn{13}{|l|}{{ \hspace{0.07\linewidth} same MJD on IAW watch web site
(as for example 48325.0 and 48325.1) }}\\
\multicolumn{13}{|l|}{{\hspace{0.05\linewidth}(2) File names differs according
to the observatory/monitoring campain:}}\\
\multicolumn{13}{|l|}{{ \hspace{0.07\linewidth} PTG* : Crimean Astrophysical
Observatory data }}\\
\multicolumn{13}{|l|}{{ \hspace{0.07\linewidth} n*   : AGN watch data}}\\
\multicolumn{9}{|l|}{{ \hspace{0.07\linewidth} 5*   : SAO/GHO data }}\\
\multicolumn{13}{|l|}{{ \hspace{0.07\linewidth} A*   : Asiago data }}\\
\hline \multicolumn{12}{|r|}{{Continued on next page}} \\ \hline
\endfoot
\hline 
\endlastfoot

41420.546  &  7.64  &  0.14  &  5.02  &  0.08  &  -3695.2  &  4127.0  &  -2750.7  &  2317.6  &  -2040.4  &  518.9  &  -1565.9  &  -137.8  \\
41426.531  &  8.38  &  0.16  &  4.86  &  0.09  &  -4131.3  &  4272.3  &  -3116.2  &  2012.3  &  -2277.5  &  737.1  &  -1677.0  &  131.7  \\
41446.471  &  6.10  &  0.02  &  4.33  &  0.01  &  -4065.1  &  3665.6  &  -3169.5  &  2749.3  &  -2451.0  &  379.8  &  -1911.0  &  -345.5  \\
41460.463  &  6.44  &  0.18  &  4.44  &  0.13  &  -4542.0  &  7289.4  &  -3349.8  &  2805.7  &  -2632.2  &  -227.5  &  -2212.7  &  -770.2 
\\
41484.456  &  8.85  &  0.15  &  6.10  &  0.05  &  -4265.2  &  4006.7  &  -3131.6  &  1932.5  &  -2413.4  &  355.8  &  -1933.6  &  -308.7  \\
41566.266  &  13.35  &  0.42  &  11.18  &  0.02  &  -6064.2  &  4237.7  &  -3640.7  &  2173.0  &  -2689.5  &  242.6  &  -2153.1  &  -418.1 
\\
41681.662  &  14.71  &  0.73  &  11.96  &  0.40  &  -5332.9  &  3892.3  &  -3795.3  &  2435.8  &  -2904.6  &  384.4  &  -2368.8  &  -516.2 
\\
41682.658  &  14.33  &  0.05  &  12.77  &  0.00  &  -4086.0  &  4141.3  &  -3136.8  &  2986.7  &  -2482.4  &  29.4  &  -2065.2  &  -630.5 
\\
\end{longtable}
\end{tiny}
\end{center}

\begin{table}[h!]
 \centering
 \small
  \caption{Results from the periodicity analysis on
  light curves of continuum at  5100\AA\, H$\beta$ fluxes,
  and radial velocity curves of half widths of broad H$\beta$
  emission line measured at blue and red side of 25\%, 50\%, 75\%
  and 90\% of maximum intensity. Note: 
 In the table are given period and
  the formal false alarm probability (P-value) for
  Lomb-Scargle periodogram  of 80 day binned
  data series, where we present longest significant periods.}
 
\label{LStbl}
\end{table}

\eject

\appendix{
\section{Appendix: [\ion{O}{3}]  calibration of slit spectra}

\begin{figure}

\includegraphics[width=.99\columnwidth]{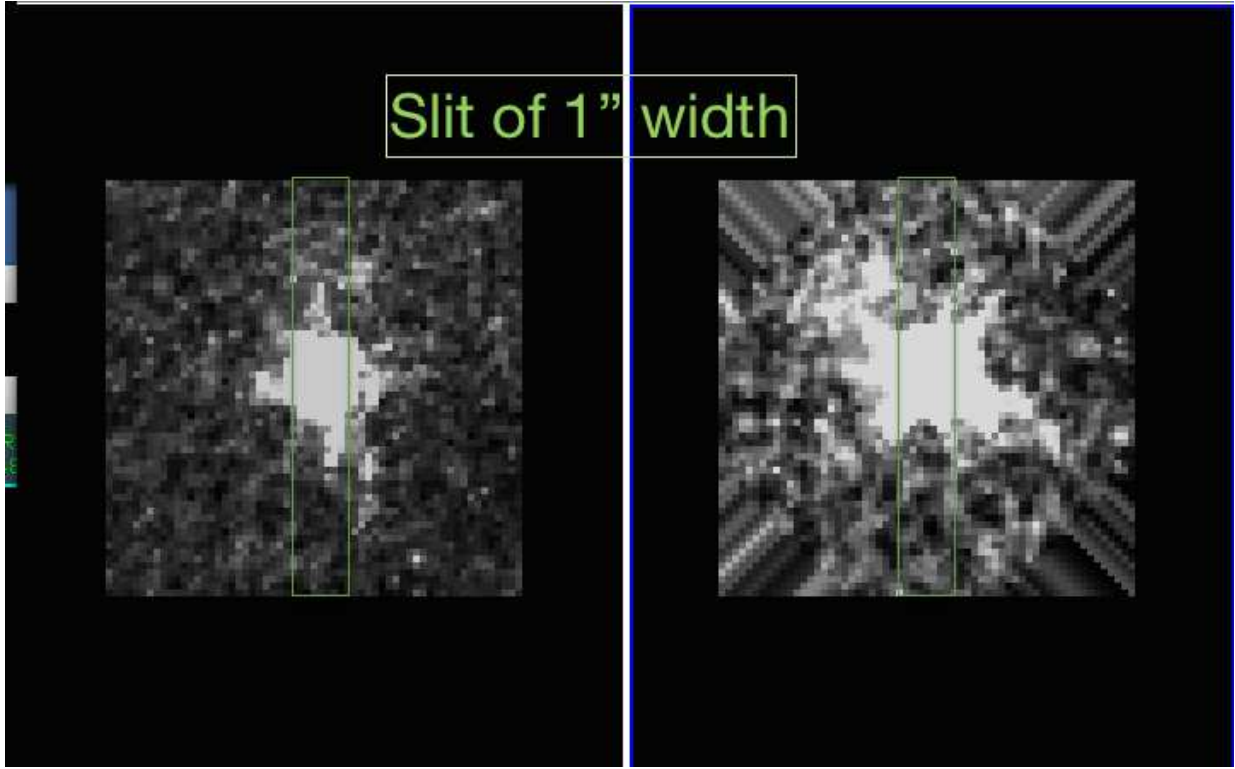}

\centering
\caption{Slit size and {position angle} influence on flux calibration using  [\ion{O}{3}]
emission {(see Table \ref{tab:oiii}). {Left panel: PA=0, Right panel: same image rotated clockwise by 45 degrees.}}
}\label{slit}
\end{figure}

NGC 5548 hosts a very compact narrow line region (NLR) structure
\citep[see][]{Kramer98, Schmitt03, Peterson2013}.
The [\ion{O}{3}] emission is so tightly concentrated that there is no position angle (PA)
effect  if the slit is wide enough, as can be seen in \cite{Schmitt03}.  For a {1 -- 2" slit the} effect is small but
appreciable{. It is about 4\% for a 1" slit, as can be seen} by comparing the total flux within the slit
at position {angles $0^{o}$ and $-45^{o}$.}
{This is shown in column 1 of Table \ref{tab:oiii}}, where the total [\ion{O}{3}] flux within the slit
$F_\mathrm{tot}$\ ({column 3}) is compared to the unresolved core flux within the slit
$F_\mathrm{slit}$\ ({column 4}). The measurements were carried out on a narrow band
WFPC2 image with angular resolution {0.1"} (see Fig.\@ \ref{slit}), available in
digital format at {NED}\footnote{http://ned.ipac.caltech.edu/img/2003ApJS..148..327S/
 NGC$\_$5548:I:{\ion{O}{3}}:sda2003.fits.gz}
 and originally published by \citet{Schmitt03}.
The total emission measured on the image is {$4.30 \times 10^{-13}$}, assuming {zero} background in the {image.
The average background is actually slightly less than zero ($\thickapprox -5 \times 10^{-17}$) but it is difficult to estimate it
accurately.}

\begin{table}
 \centering
 \small
  \caption{Slit size and {position angle} influence on flux calibration using  [\ion{O}{3}]
emission.  \label{tab:oiii}}
  \begin{tabular}{@{}lccl@{}}
  \hline
 PA     & Col. range  &    $F_\mathrm{tot}$         & $F_\mathrm{slit}$   \\
 \hline\\
\multicolumn{4}{c}{2    arcsec  virtual slit}\\
\\
0       &       23-42   &       4.28E-13        &       4.01E-13              \\
-45     &       23-42   &       4.19E-13        &       4.12E-13              \\
\\
\multicolumn{4}{c}{1    arcsec  virtual slit}                   \\
\\
0       &       28-37   &       4.03E-13        &       3.86E-13          \\
-45     &       28-37   &       3.90E-13        &       3.85E-13          \\
\hline
\end{tabular}
\end{table}
}



\newpage

\begin{thebibliography}



\bibitem[Anderson (1971)]{Anderson71} {Anderson, K., S.,1971, \apj, 169, 449}




\bibitem[Arav et
al.(2015)]{Arav2015} Arav, N., Chamberlain, C., Kriss, G.~A., et al.\ 2015, \aap, 577, A37





\bibitem[Armitage et al.(1996)]{Armitage1996} Armitage, P.~J.,
Zurek, W.~H., \& Davies, M.~B.\ 1996, \apj, 470, 237


\bibitem[Begelman et al.(1980)]{Beg80}  Begelman, M. C.,
Blandford, R. D., \& Rees, M. J.\, 1980,  {\nat},  287, 307


\bibitem[Bentz et al.(2007)]{Bentz07} Bentz, M. C., Denney,
K.~D., Cackett, E.~M., et al.\ 2007, \apj, 662, 205

\bibitem[Bentz et al.(2009)]{Bentz2009} Bentz, M.~C., Walsh,
J.~L., Barth, A.~J., et al.\ 2009, \apj, 705, 199

\bibitem[Bentz et al.(2013)]{Bentz13} Bentz, M.~C., Denney,
K.~D., Grier, C.~J., et al.\ 2013, \apj, 767, 149

\bibitem[Bentz
\& Katz(2015)]{bentzkatz15} Bentz, M.~C., \& Katz, S.\ 2015, \pasp, 127, 67


\bibitem[Bogdanovi\'c et al.(2008)]{Bog08} Bogdanovi\'c, T., S., Britton D.,
Sigurdsson, S., Eracleous, M.,
2008, \apjs, {174} , 455

\bibitem[Bogdanovi\'c et al. (2009)]{Bog09} Bogdanovi{\'c}, T., Eracleous, M., \& Sigurdsson, S.\ 2009, \nar, 53, 113

\bibitem[Bogdanovi\'c (2015)]{Bog15} Bogdanovi{\'c}, T.\ 2015, Astrophysics and Space Science Proceedings, 40, 103





\bibitem[Bon et al.(2012)]{Bon12} Bon, E., Jovanovi{\'c}, P.,
Marziani, P. et al. 2012, {\apj}, 759, 118


\bibitem[Bon et al.(2014)]{Bon2014} Bon, N., Popovi{\'c},
L. {\v C}., \& Bon, E.\ 2014, Advances in Space Research, 54, 1389




\bibitem[Bon et al. (2016)]{Bon16} Bon et al. 2016. in preparation.



\bibitem[Bouchard et al.(2010)]{Bouch10}
Bouchard, A., Prugniel, P., Koleva, M., Sharina, M., 2010, A\&A, 513, 54



\bibitem[Chen et al.(2011)]{Chen2011} Chen, X., Sesana, A., Madau, P., \& Liu, F.~K.\ 2011, \apj, 729, 13



\bibitem[Chakrabarti
\& Wiita(1993)]{Chakrabarti1993} Chakrabarti, S.~K., \& Wiita, P.~J.\ 1993, \apj, 411, 602

\bibitem[Chakrabarti
\& Wiita(1994)]{Chakrabarti94} Chakrabarti, S.~K., \& Wiita, P.~J.\ 1994, \apj, 434, 518






\bibitem[{Cherepashchuk \& Lyutyi}(1973)]{CherepashchukLyutyi73} {Cherepashchuk, A.~M., \& Lyutyi, V.~M.\ 	1973, \aplett, 13, 165}

\bibitem[Clavel et al.(1991)]{Clavel91} {Clavel, J., Reichert,
G.~A., Alloin, D., et al.\ 1991, \apj, 366, 64}


	

	
\bibitem[Cuadra et al.(2009)]{Cuadra09} Cuadra, J., Armitage,
P.~J., Alexander, R.~D., \& Begelman, M.~C.\ 2009, \mnras, 393, 1423


\bibitem[Czerny et al.(1999)]{Czerny99} Czerny, B.,
Schwarzenberg-Czerny, A., \& Loska, Z.\ 1999, \mnras, 303, 148


\bibitem[Czerny(2006)]{Czerny2006} Czerny, B.\ 2006, Astronomical
Society of the Pacific Conference Series, 360, 265
	
\bibitem[Czerny
\& Hryniewicz(2011)]{Czerny2011} Czerny, B., \& Hryniewicz, K.\ 2011, \aap, 525, L8
	
	
\bibitem[De Paolis et
al.(2003)]{DePaolis2003} De Paolis, F., Ingrosso, G., Nucita, A.~A., \& Zakharov, A.~F.\ 2003, \aap, 410, 741

\bibitem[Denney et al.(2009)]{Denney09} {Denney, K.~D., Peterson,
B.~M., Pogge, R.~W., et al.\ 2009, \apjl, 704, L80 }

\bibitem[Denney (2010)]{Denney10} Denney K., D., 2010, ProQuest Dissertations
And
Theses; Thesis (Ph.D.)--The Ohio State University, Vol.: 71-11, 101


\bibitem[Denney et al.(2014)]{Denney14} Denney, K.~D., De Rosa,
G., Croxall, K., et al.\ 2014, \apj, 796, 134

\bibitem[De Rosa et al.(2015)]{DeRosa15} De Rosa, G., Peterson,
B.~M., Ely, J., et al.\ 2015, \apj, 806, 128

\bibitem[Deuch (1966)]{Deich66} Deuch, A.~N. 1966, IAU Symposium 29

\bibitem[Dibai et al.(1968)]{Dibai68} Dibai, {\'E}.~A., Esipov,
V.~F., \& Pronik, V.~I.\ 1968, \sovast, 11, 553

\bibitem[Dietrich et al.(1993)]{Dietrich93} {Dietrich, M.,
Kollatschny, W., Peterson, B.~M., et al.\ 1993, \apj, 408, 416}
optical emission lines of NGC 5548

\bibitem[Dietrich et
al.(2001)]{Dietrich01} {Dietrich, M., Bender, C.~F., Bergmann, D.~J., et al.\ 2001, \aap, 371, 79 }







\bibitem[Du et al.(2015)]{Du15} Du, P., Hu, C., Lu, K.-X.,
et al.\ 2015, \apj, 806, 22





\bibitem[Eckart
\& Genzel(1997)]{Eckart97} Eckart, A., \& Genzel, R.\ 1997, \mnras, 284, 576


\bibitem[Edelson et al.(2015)]{Edelson2015} Edelson, R., Gelbord,
J.~M., Horne, K., et al.\ 2015, \apj, 806, 129

\bibitem[Elitzur
\& Shlosman(2006)]{Elitzur2006} Elitzur, M., \& Shlosman, I.\ 2006, \apjl, 648, L101

\bibitem[Elitzur(2008)]{Elitzur2008} Elitzur, M.\ 2008, \nar, 52,
274

{
\bibitem[Elitzur \& Netzer(2016)]{EltNetz2016} Elitzur, M., \& Netzer, H.\ 2016, \mnras, 459, 585
}

\bibitem[Elvis(2000)]{Elvis2000} Elvis, M.\ 2000, \apj, 545, 63


\bibitem[Eracleous
\& Halpern(2003)]{Eracleous2003} Eracleous, M., \& Halpern, J.~P.\ 2003, \apj, 599, 886



\bibitem[Eracleous et al.(2012)] {erac2012} Eracleous, M., Boroson, T., A.,
Halpern, J., P., Liu, J., 2012, ApJS, 201, 23


\bibitem[Fan et al.(1998)]{Fan1998} Fan, J.~H., Xie, G.~Z.,
Pecontal, E., Pecontal, A., \& Copin, Y.\ 1998, \apj, 507, 173

{
\bibitem[Fausnaugh et al.(2016)]{Fausnaugh2015} Fausnaugh, M.~M., Denney, K.~D., Barth, A.~J., et al.\ 2016, \apj, 821, 56
}

\bibitem[Flohic
\& Eracleous(2008)]{Flohic08} Flohic, H.~M.~L.~G., \& Eracleous, M.\ 2008, \apj, 686, 138



\bibitem[Gaskell(1983)]{Gaskell83} Gaskell, C. M.\ 1983, Liege
International Astrophysical Colloquia, 24, 473


\bibitem[{Gaskell}(1988)]{Gaskell88} Gaskell, C.~M. 1988, \apj, 325, 114



\bibitem[Gaskell(2008)]{Gaskell2008} Gaskell, C.~M.\ 2008, Revista Mexicana de Astronomia y Astrofisica Conference Series, 32, 1


\bibitem[Gaskell (2009)] {Gask09} Gaskell, C. M.,2009,  {\nar},  55, 140


\bibitem[Gaskell \& Klimek (2003)] {Gask03} Gaskell, C. M. \& Klimek, E. S.,
2003  {Astronomical and Astrophysical Transactions}, 22, 661
	
\bibitem[{Gaskell \& Sparke} (1986)]{GaskellSparke86} {Gaskell, C.~M. \& Sparke, L.~S.\ 1986, \apj, 305, 175}
	
	


	
	
\bibitem[Gezari et al.(2007)]{Gezari07} Gezari, S., Halpern,
J. P., \& Eracleous, M.\ 2007, \apjs, 169, 167

\bibitem[Gezari et al.(2012)]{gezarietal12} Gezari, S., Chornock,
R., Rest, A., et al.\ 2012, \nat, 485, 217


\bibitem[Ghez et al.(1998)]{Ghez98} Ghez, A.~M., Klein, B.~L.,
Morris, M., \& Becklin, E.~E.\ 1998, \apj, 509, 678

\bibitem[Gillessen et al.(2009)]{Gillessen09} Gillessen, S.,
Eisenhauer, F., Trippe, S., et al.\ 2009, \apj, 692, 1075






\bibitem[Goosmann et al.(2014)]{Goosmann2014} Goosmann, R.~W.,
Gaskell, C.~M., \& Marin, F.\ 2014, Advances in Space Research, 54, 1341

\bibitem[Graham et al.(2015a)]{Graham15a} Graham, M. J.,
Djorgovski, S. G., Stern, D., et al.\ 2015, \nat, 518, 74

\bibitem[Graham et al.(2015b)]{Graham15b} Graham, M. J.,
Djorgovski, S. G., Stern, D., et al.\ 2015, \mnras, 453, 1562



\bibitem[Guo et al.(2006)]{Guo2006} Guo, D., Tao, J.,
\& Qian, B.\ 2006, \pasj, 58, 503




\bibitem[Guo et al.(2014)]{Guo2014} Guo, D.-F., Hu, S.-M., Tao,
J., et al.\ 2014, Research in Astronomy and Astrophysics, 14, 923

\bibitem[Hayasaki et al.(2008)]{Hayasaki08}Hayasaki, K., Mineshige, S., Ho,
Luis C. 2008, \apj, 682, 1134

\bibitem[Hayasaki et al.(2015)]{Hayasaki2015} Hayasaki, K., Stone,
N.~C., \& Loeb, A.\ 2015, arXiv:1501.05207

\bibitem[Hills(1975)]{hills75} Hills, J.~G.\ 1975, \nat, 254,
295

\bibitem[Hu et al.(2008)]{Hu2008} Hu, C., Wang, J.-M., Ho,
L.~C., et al.\ 2008, \apjl, 683, L115

\bibitem[Ivanov et al.(2005)]{Ivanov2005} Ivanov, P.~B., Polnarev, A.~G., \& Saha, P.\ 2005, \mnras, 358, 1361







\bibitem[Jovanovi{\'c} et al.(2010)]{Jovanovic10} Jovanovi{\'c},
P., Popovi{\'c}, L.~{\v C}., Stalevski, M.,
\& Shapovalova, A.~I.\ 2010, \apj, 718, 168


\bibitem[Kaspi et al.(2000)]{Kaspi2000} Kaspi, S., Smith, P.~S.,
Netzer, H., et al.\ 2000, \apj, 533, 631



\bibitem[Kaastra et al.(2014)]{Kaastra2014} Kaastra, J. S., Kriss, G. A.,
Cappi, M., et al., 2014, Science, 345, 64




\bibitem[Karas \& {\v S}ubr(2007)]{KarasSubr2007} Karas, V., \& {\v S}ubr, L.\ 2007, \aap, 470, 11



\bibitem[Katz (1997)]{Katz97} Katz, J.~I.\,  (1997) {\apj}, 478, 527



{
\bibitem[Kieffer \& Bogdanovi{\'c}(2016)]{Bogdanovic2016} Kieffer, T.~F., \& Bogdanovi{\'c}, T.\ 2016, \apj, 823, 155
}

\bibitem[Koleva et al.(2008)]{Kol08}
Koleva, M., Prugniel, P., Ocvirk, P., Le Borgne, D., Soubiran, C., MNRAS, 385,
1998

\bibitem[Koleva et al.(2009)]{Kol09} Koleva, M., Prugniel, P., Bouchard, A.,
\& Wu, Y.\ 2009, \aap, 501, 1269

\bibitem[Koleva et al.(2011)]{Kol11}
Koleva, M., Prugniel, P., de Rijcke, S., Zeilinger, W. W., 2011, MNRAS, 417,
1643

 \bibitem[Koleva et al.(2013)]{Kol13}
Koleva, M., Bouchard, A., Prugniel, P., de Rijcke, S., Vauglin, I., 2013, MNRAS,
428, 2949

\bibitem[Kollatschny
\& Zetzl(2013a)]{Kollatschny13a} Kollatschny, W., \& Zetzl, M.\ 2013, \aap,
551,
L6


\bibitem[Komossa \& Bade(1999)]{Komossa99} Komossa, S., \& Bade, N.\ 1999, \aap, 343, 775



\bibitem[Komossa (2006)]{Komossa06} Komossa S., 2006, Mem. Soc. Astron. Ital., 77,
733


\bibitem[Komossa et al.(2008)]{Komossa2008} Komossa, S., Xu, D.,
Zhou, H., Storchi-Bergmann, T., \& Binette, L.\ 2008, \apj, 680, 926

\bibitem[Komossa(2015)]{Komossa15} Komossa, S.\ 2015, Journal of
High Energy Astrophysics, 7, 148



\bibitem[Komossa et al.(2016)]{KomossaZensus2015} Komossa, S., \& Zensus, J.~A.\ 2016, IAU Symposium, 312, 13




\bibitem[Koratkar
\& Gaskell(1991)]{Koratkar91} {Koratkar, A.~P., \& Gaskell, C.~M.\ 1991, \apj, 375, 85}


\bibitem[Korista et al.(1995)]{Korista95} {Korista, K.~T., Alloin,
D., Barr, P., et al.\ 1995, \apjs, 97, 285}
ground-based study of NGC 5548



\bibitem[Kova{\v c}evi{\'c} et al.(2010)]{Kovacevic2010} Kova{\v
c}evi{\'c}, J., Popovi{\'c}, L.~{\v C}.,
\& Dimitrijevi{\'c}, M.~S.\ 2010, \apjs, 189, 15


\bibitem[Koshida et al.(2014)]{koshidaetal14} Koshida, S., Minezaki,
T., Yoshii, Y., et al.\ 2014, \apj, 788, 159


\bibitem[Kramer et al.(1998)] {Kramer98} Kraemer, S.~B., Crenshaw, D.~M., Filippenko, A.~V., \& Peterson, B.~M.\ 1998, \apj, 499, 719 


\bibitem[{{Kun} et~al.}(2014)]{Kun2014}
{Kun} E.,  {Gab{\'a}nyi} K.~{\'E}.,  {Karouzos} M.,  {Britzen} S.,    {Gergely}
  L.~{\'A}.,  2014, MNRAS, 445, 1370

\bibitem[Landt et al.(2015)]{Landt15} Landt, H., Ward, M.~J.,
Steenbrugge, K.~C., \& Ferland, G.~J.\ 2015, \mnras, 454, 3688


\bibitem[Lawson \& Hanson(1995)]{LH95}
Lawson, C. L. \& Hanson, R. J., Solving Least Squares Problems,
Classics in Applied Mathematics No. 15
(Philadelphia, Penn.: SIAM), 1995


\bibitem[{Le Borgne et al.}(2004)]{LeBorgne04}
{Le Borgne D., Rocca-Volmerange B., Prugniel P., Lancon A., Fioc M., Soubiran
C., 2004, A\&A, 425, 881}

\bibitem[Lehto
\& Valtonen (1996)]{Lehto1996} Lehto, H.~J., \& Valtonen, M.~J.\ 1996, \apj, 460, 207

\bibitem[Lewis et al.(2010)] {Lewis10} Lewis, K. T., Eracleous, M. \&
Storchi-Bergmann, T. 2010, {\apjs},  {187}, 416



{
\bibitem[Li et al.(2016)]{Li2016} Li, Y.-R., Wang, J.-M., Ho, L.~C., et al.\ 2016, \apj, 822, 4
}




\bibitem[Liu et al.(2009)]{liuetal09} Liu, F.~K., Li, S.,
\& Chen, X.\ 2009, \apjl, 706, L133


\bibitem[Liu et al.(2015)]{Liu15} Liu, T., Gezari, S.,
Heinis, S., et al.\ 2015, \apjl, 803, L16

\bibitem[Liu et al.(2016)]{Liu16} Liu, J., Eracleous, M.,
\& Halpern, J.~P.\ 2016, \apj, 817, 42

\bibitem[Lomb (1976)]{lomb76} Lomb, N. R.,  1976, \apss, {39}, 447

\bibitem[Lyutyi(1973)]{Lyutyi73} {Lyutyi, V.~M.\ 1973,
Astronomicheskij Tsirkulyar, 777, 1}


\bibitem[Marquardt(1963)]{Marquardt63}
Marquardt, D. W., 1963, SIAM, 11, 431

\bibitem[MacFadyen
\& Milosavljevi{\'c}(2008)]{MacFadyen08} MacFadyen, A.~I., \& Milosavljevi{\'c}, M.\ 2008, \apj, 672, 83







\bibitem[Maoz et al.(1994)]{Maoz94} Maoz, D., Smith, P.~S., Jannuzi, B.~T., Kaspi, S., \& Netzer, H.\ 1994, \apj, 421, 34



\bibitem[Marziani et
al.(2016)]{Marziani2016} Marziani, P., Sulentic, J.~W., Stirpe, G.~M., et al.\ 2016, \apss, 361, 3

{
\bibitem[Mehdipour et
al.(2015)]{Mehdipour2015} Mehdipour, M., Kaastra, J.~S., Kriss, G.~A., et al.\ 2015, \aap, 575, A22
}

{
\bibitem[Mehdipour et al.(2016)]{Mehdipour2016} Mehdipour, M., Kaastra, J.~S., Kriss, G.~A., et al.\ 2016, \aap, 588, A139
}

\bibitem[Meyer et al.(2012)]{Meyer2012} Meyer, L., Ghez, A.~M.,
Sch{\"o}del, R., et al.\ 2012, Science, 338, 84


\bibitem[Merritt \& Milosavljevi\'c (2005)] {MM05} Merritt, D. and
Milosavljevi\'c, M., 2005, Living Reviews in Relativity, 8, 8


\bibitem[Milosavljevi\'c \& Merritt (2001)] {MM2001}  Milosavljevi\'c, M. and
Merritt, D., 2001, {\apj}, {563}, 34



\bibitem[Magorrian \& Tremaine(1999)]{MagorrianTremaine1999} Magorrian, J., \& Tremaine, S.\ 1999, \mnras, 309, 447



{
\bibitem[Nandra et al.(1997)]{Nandra97} Nandra, K., George, I.~M., Mushotzky,
R.~F., Turner, T.~J., \& Yaqoob, T.\ 1997, \apj, 476, 70
}


\bibitem[Netzer et al.(1990)]{Netzer90} {Netzer, H., Maoz, D.,
Laor, A., et al.\ 1990, \apj, 353, 108}


\bibitem[Netzer \& Peterson(1997)]{Netzer97} Netzer, H., \& Peterson, B.~M.\ 1997, Astronomical Time Series, 218, 85

\bibitem[Netzer
\& Marziani(2010)]{NetzerMarz10} Netzer, H., \& Marziani, P.\ 2010, \apj, 724, 318


\bibitem[{{Netzer}(2013)}]{Netzer2013}
{Netzer}, H. 2013, {The Physics and Evolution of Active Galactic Nuclei}, Cambridge, UK: Cambridge
University Press


\bibitem[Netzer(2015)]{Netzer15} Netzer, H.\ 2015, \araa, 53, 365

{
\bibitem[Nikolajuk et al.(2004)]{Nikolajuk2004} Nikolajuk, M.,
Papadakis, I.~E., \& Czerny, B.\ 2004, \mnras, 350, L26
}

{
\bibitem[Nguyen \& Bogdanovic(2016)]{BogNguyen2016} Nguyen, K., \& Bogdanovic, T.\ 2016, arXiv:1605.09389
}


\bibitem[Oknyanskij(1978)]{Oknyanskij78} Oknyanskij, V.~L.\ 1978, Peremennye Zvezdy, 21, 71

\bibitem[Oknyanskij
\& Lyuty(2007)]{Oknyanskij2007} Oknyanskij, V., \& Lyuty, V.\ 2007, Peremennye Zvezdy Prilozhenie, 7, 28






\bibitem[Peterson
\& Gaskell(1986)]{PetersonGaskell86} {Peterson, B.~M., \& Gaskell, C.~M.\ 1986, \aj, 92, 552}

\bibitem[Peterson et al.(1987)]{Peterson87} Peterson, B. M., Korista, K. T.,
Cota, S. A., 1987, \apjl, 312, 1

\bibitem[Peterson et al.(1991)]{Peterson91} {Peterson, B.~M.,
Balonek, T.~J., Barker, E.~S., et al.\ 1991, \apj, 368, 119 }
5548 at optical wavelengths

\bibitem[Peterson et al.(1992)]{Peterson92} {Peterson, B.~M.,
Alloin, D., Axon, D., et al.\ 1992, \apj, 392, 470 }
5548 at optical wavelengths

\bibitem[Peterson(1997)]{Peterson97} Peterson, B.~M.\ 1997, An introduction to active galactic nuclei, Publisher: Cambridge, New York
Cambridge University Press, 1997 Physical description xvi, 238 p.~ISBN 0521473489


\bibitem[Peterson et al.(1999)]{Peterson99} Peterson, B.~M.,
Barth, A.~J., Berlind, P., et al.\ 1999, \apj, 510, 659

\bibitem[Peterson et al.(2002)]{Peterson02}
Peterson, B. M., Berlind, P., Bertram, R., et al. 2002, \apj, 581, 197


\bibitem[Peterson et al.(2013)]{Peterson2013} Peterson, B. M.,
Denney, K. D., De Rosa, G., et al.\ 2013, \apj, 779, 109

\bibitem[Pihajoki(2016)]{Pihajoki2016} Pihajoki, P.\ 2016, \mnras,
457, 1145


\bibitem[Plavchan et al.(2008)]{Plavchan2008} Plavchan, P., Jura,
M., Kirkpatrick, J.~D., Cutri, R.~M.,
\& Gallagher, S.~C.\ 2008, \apjs, 175, 191




\bibitem[Popovi{\'c} et al.(2008)]{Popovic2008} Popovi{\'c}, L.~{\v
C}., Shapovalova, A.~I., Chavushyan, V.~H., et al.\ 2008, \pasj, 60, 1

\bibitem[Popovi\'c (2012)] {Pop2012} Popovi\'c, L, \v C., 2012, {\nar}, {56}, 74



\bibitem[Pringle (1996)] {Pring96} Pringle, J. E.\, 1996, {\mnras}, {281}, 357


\bibitem[Ptak \& Stoner (1973)]{Ptak73} Ptak, R., L. \& Stoner, R., E., 1973,
\apj, 179, 89

	

\bibitem[Rees(1990)]{Rees90} Rees, M.~J.\ 1990, Science, 247, 817


\bibitem[Rieger
\& Mannheim(2000)]{Rieger2000} Rieger, F.~M., \& Mannheim, K.\ 2000, \aap, 359, 948




\bibitem[Rix \& White(1992)]{RW92}
Rix, H.-W., \& White, S. D. M., 1992, MNRAS, 254, 389

{
\bibitem[Rokaki et al.(1993)]{Rokaki93} Rokaki, E., Collin-Souffrin, S., \& Magnan, C.\ 1993, \aap, 272, 8
}


\bibitem[Scargle (1982)]{scar82} Scargle, J. D., 1982, \apj  {263}, 835

\bibitem[Schmitt et al.(2003)]{Schmitt03} Schmitt, H.~R., Donley,
J.~L., Antonucci, R.~R.~J., Hutchings, J.~B.,
\& Kinney, A.~L.\ 2003, \apjs, 148, 327


\bibitem[Sergeev (1992)]{Sergeev92} Sergeev  S. G., 1992, Ap\&SS, 197, 77


\bibitem[Sergeev et al.(2007)]{Sergeev07} Sergeev, S. G., Doroshenko, V. T.,
Dzyuba, S. A., et al. 2007, \apj, 668,
708


\bibitem[Shapovalova et al.(2004)]{Shap04} Shapovalova, A., I., Doroshenko, V.,
T., Bochkarev, N., G., et al., 2004, {\aap}, 422, 925

\bibitem[Shapovalova et al.(2006)]{Shap06} Shapovalova, A. I.,
Burenkov, A. N., Borisov, N., et al., 2006, Astronomical Society of the
Pacific Conference Series, 360, 239

\bibitem[Shapovalova et al.(2008)]{shap08} Shapovalova, A. I., Popovi{\'c},
L. {\v C}., Collin, S., et al. 2008,   \aap,  {486}, 99

{
\bibitem[Shapovalova et al.(2016)]{Shap16} Shapovalova, A.~I., Popovi{\'c}, L.~{\v C}., Chavushyan, V.~H., et al.\ 2016, \apjs,
222, 25
}





\bibitem[Shen \& Loeb (2010)]{Shen10} Shen, Y. \& Loeb, A. 2010, \apj, {725},
249





\bibitem[Sillanpaa et al.(1988)]{Sillanpaa88} Sillanpaa, A.,
Haarala, S., Valtonen, M.~J., Sundelius, B.,
\& Byrd, G.~G.\ 1988, \apj, 325, 628



\bibitem[Slavcheva-Mihova
\& Mihov(2011)]{Slavcheva2011} Slavcheva-Mihova, L., \& Mihov,
B.\ 2011, \aap, 526, A43




\bibitem[Smailagi{\'c}
\& Bon(2015)]{SmaBon2015} Smailagi{\'c}, M., \& Bon, E.\ 2015, Journal of Astrophysics and Astronomy, 36, 513

\bibitem[Smith
\& Hoffleit(1963)]{SmitHoff1963} Smith, H.~J., \& Hoffleit, D.\ 1963, \aj, 68, 292





{
\bibitem[Stalevski et al.(2016)]{Stalevski16} Stalevski, M., Ricci, C., Ueda, Y., et al.\ 2016, \mnras, 458, 2288
}

\bibitem[Steenbrugge et
al.(2005)]{Steenbrugge2005} Steenbrugge, K.~C., Kaastra, J.~S.,
Crenshaw, D.~M., et al.\ 2005, \aap, 434, 569


\bibitem[Stellingwerf(1978)]{Stellingwerf1978} Stellingwerf, R.~F.\ 1978, \apj, 224, 953 


\bibitem[Sudou et al.(2003)]{Sudou2003} Sudou, H., Iguchi, S.,
Murata, Y., \& Taniguchi, Y.\ 2003, Science, 300, 1263


\bibitem[Syer et al.(1991)]{Syer1991} Syer, D., Clarke, C.~J.,
\& Rees, M.~J.\ 1991, \mnras, 250, 505



\bibitem[{\v S}ubr
\& Karas(1999)]{Subr1999} {\v S}ubr, L., \& Karas, V.\ 1999, \aap, 352, 452

\bibitem[Torres et
al.(2010)]{Torres2010} Torres, G., Andersen, J., \& Gim{\'e}nez, A.\ 2010, \aapr, 18, 67

\bibitem[Tyson et al.(1998)]{Tyson1998}
{Tyson}, J.~A.,  {Fischer}, P.,  {Guhathakurta}, P.,  {McIlroy}, P.,  {Wenk}, R.,
  {Huchra}, J.,  {Macri}, L.,  {Neuschaefer}, L.,  {Sarajedini}, V.,  {Glazebrook},
  K.,  {Ratnatunga}, K.,    {Griffiths}, R.,  1998, AJ, 116, 102


\bibitem[Ulrich (1972)]{Ulrich72} {Ulrich, M. H., 1972, \apj, 174, 483}



\bibitem[Uttley et al.(2003)]{Uttley2003} Uttley, P., Edelson, R., McHardy, I.~M., Peterson, B.~M., \& Markowitz, A.\ 2003,
\apjl, 584, L53

\bibitem[Valtonen \& Ciprini (2012)] {Valtonen12} Valtonen, M., Ciprini, S.,
2012, \memsai, 83, 219


\bibitem[Valtonen
\& Sillanp{\"a}{\"a}(2011)]{ValtonenSila2011} Valtonen, M., \& Sillanp{\"a}{\"a}, A.\ 2011, Acta Polytechnica, 51, 76

\bibitem[Valtaoja et al.(2000)]{Valtaoja2000} Valtaoja, E.,
Ter{\"a}sranta, H., Tornikoski, M., et al.\ 2000, \apj, 531, 744



\bibitem[van der Marel (1994)]{vanM94}
van der Marel, R. P., 1994, \mnras, 270, 271




\bibitem[Vaughan et al.(2016)]{Vaughan2016} Vaughan, S., Uttley, P., Markowitz, A.~G., et al.\ 2016, \mnras, 461, 3145

\bibitem[Westman et al.(2011)]{Westman2011} Westman, D.~B., MacLeod, C.~L., \& Ivezi{\'c}, {\v Z}.\ 2011, Astronomical Data Analysis
Software and Systems XX, 442, 159


\bibitem[Witzel et al.(2014)]{Witzel2014} Witzel, G., Ghez, A.~M., Morris, M.~R., et al.\ 2014, \apjl, 796, L8



\bibitem[Wu et al.(2011)]{Wu11} Wu, Y., Singh, H. P., Prugniel, P., Gupta, R., Koleva, M., 2011, \aap, 525, 71










\bibitem[Zu et al.(2011)]{Zu11} Zu, Y., Kochanek, C.~S.,
\& Peterson, B.~M.\ 2011, \apj, 735, 80

\bibitem[Zucker et al.(2006)]{Zucker2006} Zucker, S., Alexander,
T., Gillessen, S., Eisenhauer, F., \& Genzel, R.\ 2006, \apjl, 639, L21







\end{thebibliography}
\end{document}